\documentclass[11pt]{article}

\usepackage{a4wide}
\usepackage{xspace}
\newcommand{\be}{\begin{equation}}
\newcommand{\ee}{\end{equation}}

\usepackage{amsmath}
\usepackage{graphicx}
\usepackage{epsfig}
\usepackage{hyperref}
\usepackage{url}
\usepackage{amssymb}

\newcommand{\filt}{\mathrm{filt}}

\newcommand{\JD}{\mathrm{JD}}

\newcommand{\bea}{\begin{eqnarray}}
\newcommand{\eea}{\end{eqnarray}}

\newcommand{\lp}{\left(}
\newcommand{\rp}{\right)}

\newcommand{\cL}{{\cal L}}

\newcommand{\ie}{\emph{i.e.}\ }
\newcommand{\eg}{\emph{e.g.}\ }
\newcommand{\cnf}{\emph{cf.}\ }

\def\order#1{{\cal{O}}\left(#1\right)}

\newcommand{\GeV}{\,\mathrm{GeV}}
\newcommand{\TeV}{\,\mathrm{TeV}}

\newcommand{\mb}{\,\mathrm{mb}}
\newcommand{\unit}[1]{\,\mathrm{#1}}

\def\qq{\ensuremath{q\bar q}\xspace}
\def\gg{\ensuremath{gg}\xspace}

\newcommand{\ttbar}{\ensuremath{t\bar{t}}\xspace}
\newcommand{\Qa}[1]{\ensuremath{Q_{f=#1}^{w}}\xspace}
\newcommand{\Qb}[1]{\ensuremath{Q_{w=#1\sqrt{M}}^{1/f}}\xspace}
\newcommand{\rhoL}{\ensuremath{\rho_{\cal L}}\xspace}

\title{\textbf{Quantifying the performance of jet definitions\\ for
    kinematic reconstruction at the LHC}}
\author{Matteo Cacciari, Juan Rojo and Gavin P. Salam \\
  {\small \it LPTHE,}\\
 {\small \it UPMC -- Paris 6,}\\
  {\it\small
    Paris-Diderot -- Paris 7, }\\
 {\small \it CNRS UMR 7589,
    Paris, France}\\[18pt]
Gregory Soyez \\
  {\small \it Brookhaven National Laboratory, Upton, NY, USA}\\[10pt]
}
\date{}

\begin{document}

\maketitle

\vspace{-9.5cm}
 \begin{flushright}
   October 2008\\
 \end{flushright}
\vspace{7.5cm}

\begin{center}
{\bf Abstract:}
\end{center}

We present a strategy to quantify the performance of jet definitions in
kinematic reconstruction tasks. It is designed to make use exclusively
of physical observables, in contrast to previous techniques which
often used unphysical Monte Carlo partons as a reference. It is
furthermore independent of the detailed shape of the kinematic
distributions.
We analyse the performance of 5 jet algorithms over a broad range of
jet-radii, for sources of quark jets and gluon jets, spanning the
energy scales of interest at the LHC, both with and without pileup.
The results allow one to identify optimal jet definitions for the
various scenarios.
They
confirm that the use of a small jet radius ($R\simeq 0.5$)
for quark-induced jets at moderate
energy scales, $\order{100\GeV}$,  is a good
choice. However, for gluon jets and in general for TeV scales, there
are significant benefits to be had from using larger radii, up to
$R\gtrsim 1$.
This has implications for the span of jet-definitions that the LHC
experiments should provide as defaults for searches and other physics
analyses.

\clearpage

\tableofcontents

\section{Introduction}

A recurring question in jet studies is ``what is the best jet
definition for a given specific analysis''? 
One approach to answering such a question is to repeat the analysis
for a large range of jet definitions and selecting the best, whatever
this means in the context of the given analysis.
This can be rather time consuming. Furthermore, experiments may not
have easy access to a sufficiently large array of jet definitions --- for example
only a handful may be calibrated and included as standard in event
records.
It is therefore important to have advanced knowledge about the types
and the span of jet definitions that are likely to be optimal,
independently of details of specific analyses.

In this paper we investigate the question of identifying optimal jet
definitions 
with the help of characterisations of jet-finding ``quality''
that are designed to be robust and physical,  as well as reasonably
representative of a jet definition's quality for kinematic
reconstruction tasks. 
We concentrate on kinematic reconstructions (rather than more
QCD-oriented measurements, such as the inclusive-jet spectrum),
because they are a key element in a wide range of
LHC investigations, including top-quark studies and new-particle
searches. We already presented in~\cite{Buttar:2008jx} a similar, though less
systematic and extensive, investigation.

The quality of a given jet definition may depend significantly on the
process under consideration, \ie how the jets are produced.
Here we will examine both quark and gluon-induced jets, spanning a range
of energies.
They will be obtained from Monte Carlo production and decay of
fictitious narrow $Z'$ and $H$ bosons, with $Z'\to q\bar q$ and $H\to
gg$.
For each generated event we will cluster the event into jets with
about $50$ different jet definitions and determine the invariant mass
of the sum of the two hardest jets.
The distribution of invariant masses should then have a peak near
the heavy boson mass. We will take the sharpness of that peak to be
indicative of the quality of each jet definition.
By scanning a range of $Z',H$ masses we will establish this
information for a range of partonic energies.
$Z'\to q\bar q$ and $H\to g g$ events are comparatively simple,
perhaps overly so, therefore we will complement them with studies of
fully hadronic decays of $t\bar t$ events.

Our approach differs crucially from usual investigations of jet-definition
quality in that it avoids any matching to unphysical Monte Carlo
partons, whose relation to the jets can depend as much on the
details of the Monte Carlo showering algorithm (for example its
treatment of recoil) as on the jet definition.
A reflection of this is that in modern tools such as
MC@NLO~\cite{Frixione:2002ik} and POWHEG~\cite{Nason:2004rx}, which
include exact NLO corrections, the original Monte Carlo parton does
not even exist.

A further issue that we address relates to the measurement of the
sharpness of the peak. Past approaches have involved, for example,
fitting a Gaussian to the peak (see \eg \cite{fit-gaussian-1,fit-gaussian-2})
and then using its standard deviation as the quality measure. This
(and related methods) are however quite unsuited to the range of peak
shapes that arise, and we will therefore devise strategies for
measuring peak-quality independently of the precise peak shape.

The results presented in the sections below, without pileup in section
\ref{sec:results} and with pileup in section \ref{sec:results-pu}, are
complemented by an interactive web-site~\cite{quality.fastjet.fr},
which collects a far broader range of plots than can be shown here.

\section{Analysis chain}
\label{sec:gen-strat}

\subsection{Event generation}
\label{sec:ev-gen}

We consider the following processes in $pp$ collisions at $14\TeV$
centre of mass energy: $\qq \to Z' \to q\bar{q}$, as a source of quark jets
with well-defined energies, for values of $M_{Z'}$ from
$100\GeV$ to $4\TeV$; $\gg \to H \to g g$ as a source of gluons jets
(as done also by B\"uge et al. in \cite{Buttar:2008jx}), in a
similar mass range; and 
fully hadronic $t\bar t$ events with $M_t = 175\GeV$ and $M_W =
80.4\GeV$, as an example of a more complex environment.

For the $Z'$ and $H$ samples, the heavy-boson width has been
set to (a fictitious value) of less than $1\GeV$ so as to produce a
$\delta$-like peak for ideal mass reconstruction, or equivalently
so as to provide a mono-energetic source of jets.
One should be aware also that the span of masses used here does not
correspond to a physically sensible range for real Higgs or $Z'$
bosons. This is not an issue, insofar as we are only interested in the
Higgs and $Z'$ as well-defined sources of quarks and gluons in Monte
Carlo studies. To emphasise this fact, in what follows we shall refer
simply to the ``\qq'' and ``\gg'' processes.

All the samples have been generated with {\tt Pythia} 6.410
\cite{Sjostrand:2006za} with the DWT tune~\cite{Albrow:2006rt}. For
the $t\bar{t}$ samples the B mesons have been kept stable.

\subsection{Jet definitions}
\label{sec:jet-defs}

A jet definition~\cite{Buttar:2008jx} is the combination of a jet
algorithm, its parameters (\eg the radius $R$) and choice of recombination scheme. It
fully specifies a mapping from particles to jets.

We will study the following infrared and collinear-safe jet
algorithms:
\begin{enumerate}
\item the longitudinally invariant inclusive $k_t$ algorithm
  \cite{Catani:1991hj,Catani:1993hr,Ellis:1993tq}, a
  sequential-recombination algorithm whose distance measure is the relative
  transverse momentum between particles.
\item The Cambridge/Aachen (C/A) algorithm
  \cite{Dokshitzer:1997in,Wobisch:1998wt}, also
  a sequential-recombination algorithm, which uses the rapidity-azimuth
  separation between particles as its distance measure.
\item The anti-$k_t$ algorithm \cite{Cacciari:2008gp}, yet another
  sequential-recombination algorithm, with the property that it
  produces conical jets (akin to an iterative cone algorithm with
  progressive removal, such as the current CMS cone algorithm, but
  without the corresponding collinear unsafety issues).
\item SISCone \cite{Salam:2007xv}, a seedless-cone type algorithm with
  a split--merge step, whose overlap threshold is set to
  $f=0.75$.\footnote{The value of $f$ has been chosen to avoid the
    ``monster-jets''~\cite{Cacciari:2008gn} that can appear with a
    previously common default of $f=0.5$.} Additionally, we use the
  default choices of an infinite number of passes and no $p_T$-cut on
  stable cones.
\item C/A with filtering (see below).
\end{enumerate}
In each case, we will add four-momenta using $E$-scheme (4-vector)
recombination.
The algorithms all have a parameter $R$, the jet radius, which
controls the opening angle of the jets in the rapidity--azimuth plane.
Results will be quite sensitive to the choice of $R$ and we will vary
it over a suitable range for each process.

In the case of the C/A algorithm, whose clustering sequence is ordered
in rapidity-azimuth distance ($\sim$ emission angle), we will also
consider the impact of a filtering procedure~\cite{Butterworth:2008iy}
in which, subsequent to the jet finding, each jet is unclustered down
to subjets at angular scale $x_\filt R$ and one retains only the
$n_\filt$ hardest of the subjets.  We use $x_\filt = 0.5$ and
$n_\filt=2$.
Filtering is designed to limit sensitivity to the underlying event
while retaining the bulk of perturbative radiation. It is a new
technique and our scope is not to investigate it in depth (for
instance by also varying $x_\filt$ and $n_\filt$), but rather to
examine its potential beyond its original context.

All the jet algorithms have been used in the implementations and/or
plugins of the {\tt FastJet} package~\cite{Cacciari:2005hq,fastjet_web}, version
2.3, with the exception of C/A with filtering, which will be made
public in a forthcoming {\tt FastJet} release.

\subsection{Event selection and analysis}
\label{sec:ev-sel+analysis}

For each event in the \qq and \gg processes, the
reconstruction procedure is the following:
\begin{enumerate}
\item Carry out the jet finding using all final-state particles,
  taking the definition of hadron level proposed as standard in
  \cite{Buttar:2008jx}.

\item Keep only events in which the two hardest jets have $p_T \ge
  10\GeV$, $|y| \le 5$ and rapidity difference $|\Delta y| \le 1$ (the
  last of these conditions ensures that the corresponding hard partons cover a
  limited range of transverse momenta, close to $M/2$).
\item Reconstruct the invariant mass of the two hardest jets.
\end{enumerate}
For the fully hadronic $t\bar t$ process:
\begin{enumerate}
\item Carry out the jet finding as above.
\item Keep only events in which the 6 hardest jets have $p_T \ge 10\GeV$
  and $|y|\le 5$, and of which exactly two are $b$-tagged (\ie
  contain one or more $B$-hadrons).
\item Using the four non $b$-tagged jets, consider the 3 possible
  groupings into two pairs (\ie two candidate $W$-bosons). For each
  grouping, calculate the invariant mass of each pair of jets and keep
  the grouping that minimises $(M_{i_1i_2} - M_W)^2 + (M_{i_3 i_4} -
  M_W)^2$. 
\item Reconstruct the invariant masses for the two top quarks by
  pairing the $b$ and $W$ jets.  The ambiguity in the $bW$ pairing is
  resolved by taking the solution that minimises the mass difference
  between the two candidate top quarks.
\end{enumerate}
Note that we do not pass the events through any form of detector
simulation. However the choices of quality-measures that we use to
study the invariant-mass distributions will, in part, take into
account known detector resolutions.

\subsection{Figures of merit}
\label{sec:figures_of_merit}

The above procedure will give us invariant mass distributions for each
\qq (\ie $Z'$), and \gg ($H$) mass (and for $W$ and top in the $t\bar t$ sample) and for each jet
definition. We next need to establish a systematic procedure for
measuring the peak quality in each distribution.
One option would be to use a Gaussian fit to the peak.
Figure~\ref{fig:jetalgs_fits} illustrates the main difficulty with
this option, \ie that invariant mass peaks are anything but Gaussian.
One might also consider using the variance of the invariant mass
distribution. This, however, fares poorly in reflecting the quality of
the peak because of a large sensitivity to the long tails of any
distribution.
We therefore need to devise peak-quality measures that are independent
of any specific functional parametrisation, and truly reflect the
nature of the peak itself.

Our logic will be the following.  Given two peaks, if they both have
similar widths then it is the taller one that is better; if they have
similar numbers of events then it is the narrower one that is better.
These considerations lead to us to define the two possible measures:
\begin{figure}[t]
  \centering
  \includegraphics[width=0.48\textwidth]{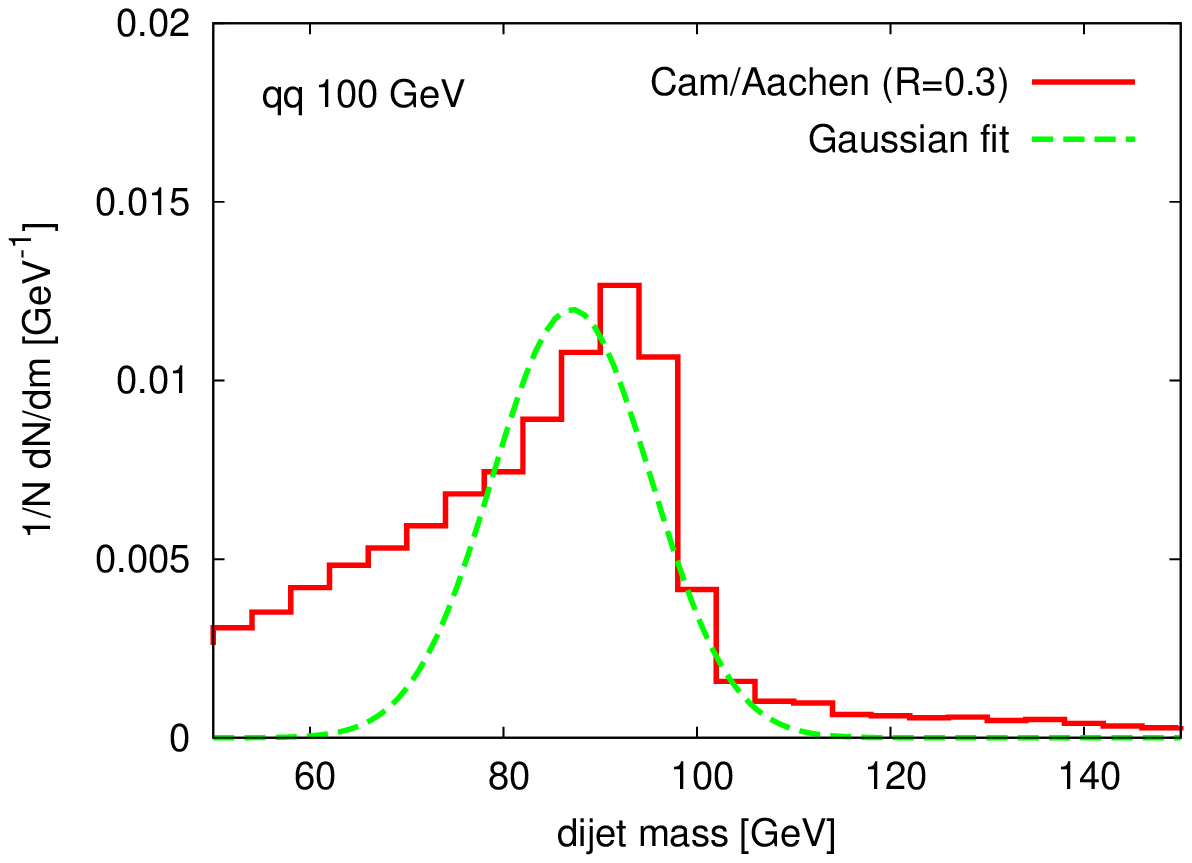}
  \includegraphics[width=0.48\textwidth]{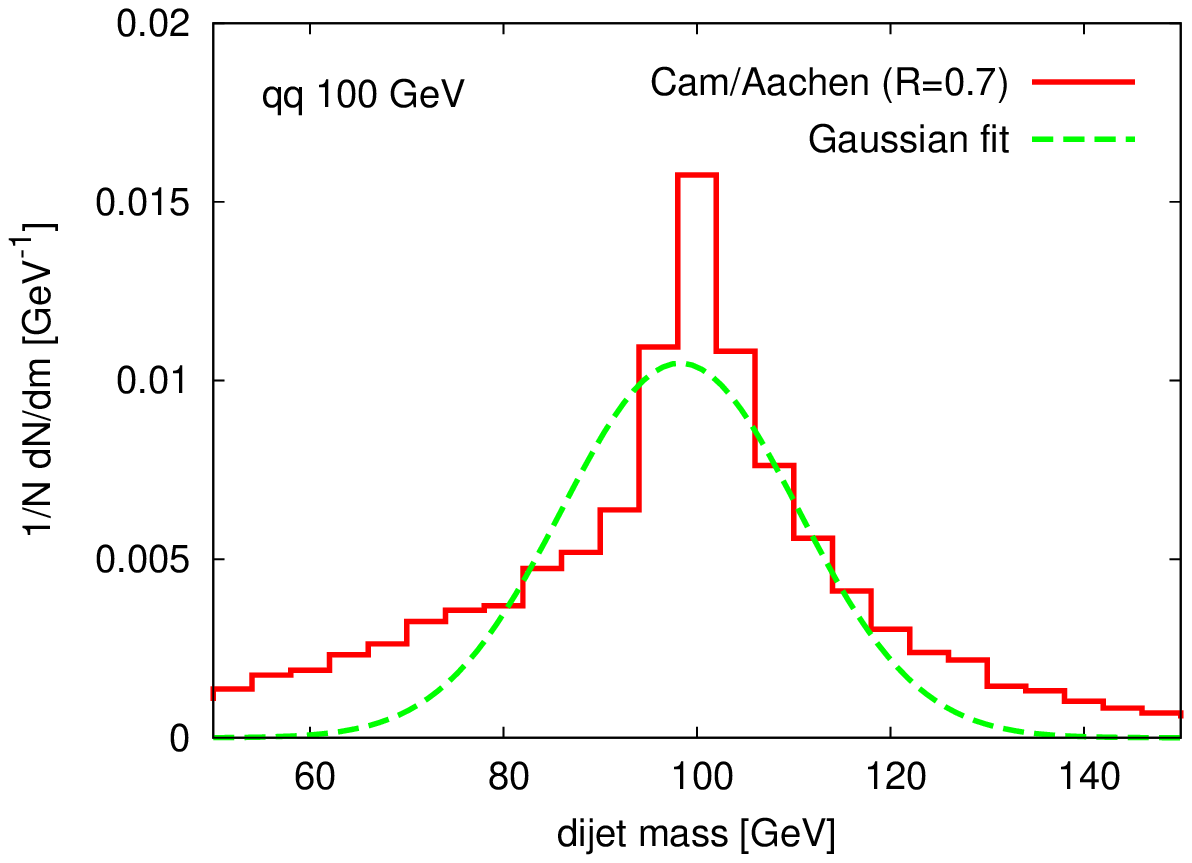}
  \caption{\small The dijet mass distribution for \qq events with
    $M_{q\bar q} = 100\GeV$, compared to a Gaussian fitted for
    reconstructed dijet masses between $75$ and $125\GeV$, for two jet
    definitions: Cambridge/Aachen with $R=0.3$ (left) and $R=0.7$
    (right).}
  \label{fig:jetalgs_fits} 
\end{figure}
\begin{enumerate}
\item {\bf $\Qa{z}$}: the width of the smallest (reconstructed)
  mass window that contains a fraction $f=z$ of the generated massive
  objects,\footnote{The number of generated
    massive objects can differ from the total number of events. For
    example if in the \ttbar samples we have $N_{\rm ev}=10^5$, the
    number of generated W bosons (and top quarks) is $N_{\rm W}=2\cdot
    10^5$.}  that is 
\begin{equation}
\label{eq:q1}
f \equiv 
 \lp \frac{{\rm \#~reco.~massive~objects~in~window~of~width~}w}{
\rm Total ~\#~generated~massive~objects}\rp  = z\,.
\end{equation}
A jet definition that is more effective in reconstructing the majority
of massive objects within a narrow mass peak gives a lower value for
$Q_{f=z}^{w}$. Therefore smaller values $Q_{f=z}^{w}$ indicate
``better'' jet definitions.

\begin{table}
\begin{center}
\begin{tabular}{|c|c|c|c|c|}
\hline
Process &  \# Gen. events  & \# Acc. events    &  Fraction acc.
vs. gen.  & $z$ in eq.~(\ref{eq:q1})\\
\hline
$Z'\to q\bar{q}$    &  50$\,$000 & $\sim$ 23$\,$000 & $\sim 0.46$ &  0.12 \\
$H \to gg$          &  50$\,$000 & $\sim$ 27$\,$000 & $\sim 0.54$ &  0.13 \\
Hadronic \ttbar     & 100$\,$000 & $\sim$ 75$\,$000 & $\sim 0.75$ &  0.18 \\
\hline
\end{tabular}
\end{center}
\caption{\small \label{tab:jetalgs_frac}  Number of
  generated events, and those accepted after event selection cuts,
  together with the fraction of generated events that this corresponds
  to, and finally 
  the value of $z$ used in eq.~(\ref{eq:q1}) (chosen to be roughly
  $1/4$ of the previous column).
}
\end{table}

Note that we normalise to the total number of generated objects rather
than the (smaller) number of objects corresponding to the events that
pass the selection cuts. This ensures that we do not favour a jet
definition for which only an anomalously small fraction of events pass
the selection cuts (as can happen in the $t\bar t$ events for large
$R$, where jets are often spuriously merged), even if the $\JD$ gives
good kinematic reconstruction on that small fraction.

The value of $z$ will be chosen, separately for each process, so that
with a typical $\JD$ the window contains about $25\%$ of the massive
objects in the events that pass the cuts. The values used for $z$ are
listed in table~\ref{tab:jetalgs_frac}.

\item {\bf $Q_{w=x\sqrt{M}}^{1/f}$}: to compute this quality measure, we
  take a window of fixed width $w$ and slide it over the mass
  distribution so as to as maximise its contents. 
  Then the figure of merit is given by
  \be
  Q_{w=x\sqrt{M}}^{1/f} \equiv
  \lp 
  \frac{{\rm Max ~\#~reco.~massive~objects~in~window~of~width~} w=x\sqrt{M} }{
    \rm Total ~\#~generated~massive~objects}\rp^{-1} \ ,
  \ee
  where the inverse has been taken so that a better jet definition leads
  to a smaller $Q_{w=x\sqrt{M}}^{1/f}$, as above.
  We set the width equal to $x\sqrt{M}$, where $M$ is the nominal
  heavy object mass and $x$ a constant to be chosen. This reflects the
  characteristic energy-dependence of resolution in hadronic
  calorimeters. We take $x = 1.25\sqrt{\mathrm{GeV}}$, a value that is
  in the ballpark of currently quoted
  resolutions for the CMS and ATLAS experiments. The reader should be
  aware that this choice is associated with a degree of arbitrariness.

\end{enumerate}

In tests of a range of possible quality measures for mass
reconstructions (including Gaussian fits, and the width at half peak
height), the above two choices have been found to be the least
sensitive to the precise shape of the reconstructed mass distribution,
and have the advantage of being independent of the binning of
the distribution.
Another encouraging feature,
which will be seen below, is that the two measures both lead to similar
conclusions on the optimal algorithms and $R$ values.

\subsection{Quantitative interpretation of figures of merit}
\label{sec:lumi-ratios}

It is useful to establish a relation between variations of the above
quality measures and the corresponding variation of integrated
luminosity needed to maintain constant significance for a signal
relative to
background.
This will allow us to quantify the importance of any differences 
between jet definitions ($\JD)$ and the potential gain to be had in using
the optimal one.
The relation will be relevant in the case in which the intrinsic width
of the physical resonance that one is trying to reconstruct is no
larger than our (narrow-resonance based) reconstructed dijet-peak ---
for a very broad physical resonance, the jet-reconstruction quality
instead becomes irrelevant.

Our relation will be valid for two background scenarios: one in which
the background is flat and independent of the jet definition; and
another in which the background is not necessarily flat, but the signal peak 
and the background shift together as one changes the jet definition (and
the second derivative of the background distribution is not too
large). For both scenarios, in a
window centred on the signal peak, the number of background events will
be proportional to the window width, and the constant of proportionality
will be independent of the jet definition.

The significance of a signal with respect to the background,
\be
\Sigma\lp \JD\rp \equiv \frac{N_{\rm signal}^\JD}{
\sqrt{N^\JD_{\rm bkgd}}} \ ,
\ee 
where $N_{\rm signal}$ and $N_{\rm bkgd}$ are respectively the number
of signal and background events, can then be rewritten as
\be
\Sigma\lp \JD\rp =  \frac{N^\JD_{\rm signal}}{
\sqrt{C\,w^\JD}} \ ,
\ee
where $w^\JD$ is the width of the window in which we count signal and
background events. The argument $\JD$ serves as a reminder that the
significance will in general depend on the jet definition, and $C$ is a
constant independent of the jet definition thanks to our assumptions
above on the structure of the background.

We can now establish the following relations between ratios of quality
measures for two jet definitions and corresponding ratios of
significance. The latter will then relate directly to ratios of
luminosities needed to achieve the same significance.

\begin{itemize}
\item in the case of \Qa{z} the number of signal events is kept fixed
and the window width depends on the jet definition. We have then
\begin{equation}
   \frac{\Sigma\lp \JD_1 \rp} {\Sigma\lp \JD_2 \rp}
  =   \left[\frac{N^{\JD_2}_{\rm bkgd}}{N^{\JD_1}_{\rm bkgd}}\right]^{1/2}
  =   \left[\frac{w^{\JD_2}}{w^{\JD_1}}\right]^{1/2}
  =  \left[\frac{Q_{f=z}^{w}\lp\JD_2\rp} {Q_{f=z}^{w}\lp\JD_1
  \rp}\right]^{1/2} \; .
\end{equation}

\item In the case of $Q_{w=x\sqrt{M}}^{1/f}(R)~$ the window width is kept
constant and it is instead the number
of signal events in the window that depends on the jet definition. Hence
\begin{equation}
   \frac{\Sigma\lp \JD_1 \rp} {\Sigma\lp \JD_2 \rp}
  =   \frac{N^{\JD_1}_{\rm signal}}{N^{\JD_2}_{\rm signal}}
  =  \frac{Q_{w=x\sqrt{M}}^{1/f}\lp\JD_2\rp} {Q_{w=x\sqrt{M}}^{1/f}\lp\JD_1
  \rp} \; .
\end{equation}
\end{itemize}

Both of these expressions are consistent with the statement that a larger
value of a quality measure indicates a worse jet definition (\ie\
the significance is smaller at fixed integrated luminosity $\cL$). This in turn implies that a larger
integrated luminosity will be needed to obtain a given fixed significance. It is
convenient to express this in terms of an effective luminosity ratio,
\begin{equation}
  \label{eq:rhol_basic_def}
  \rho_{\cal L}(\JD_2 / \JD_1) \equiv 
  \frac{{\cal L}(\text{needed with }\JD_2)}
       {{\cal L}(\text{needed with }\JD_1)} 
  = \left[
    \frac{\Sigma\lp \JD_1 \rp} {\Sigma\lp \JD_2 \rp} \right]^2 \; .
\end{equation}
Given a certain signal significance with $\JD_1$, $\rho_{\cal
  L}(\JD_2/\JD_1)$ indicates the factor more luminosity needed to
obtain the same significance with $\JD_2$.\footnote{Alternatively,
  for a fixed integrated luminosity,
  $\sqrt{\rho_{\cal L}(\JD_2/\JD_1)}$ indicates the extra factor of signal
  significance that would be gained with $\JD_1$ compared to $\JD_2$.}
The expressions for $\rhoL$ in terms of the two quality measures are
\begin{equation}
  \label{eq:rhol_minwidth}
\rho_{\cal L}(\JD_2/\JD_1) = \frac{Q_{f=z}^{w}\lp\JD_2\rp}
{Q_{f=z}^{w}\lp\JD_1 \rp} \; ,
\end{equation}
and
\begin{equation}
  \label{eq:rhol_maxfrac}
\rho_{\cal L}(\JD_2/\JD_1) = \left[\frac{Q_{w=x\sqrt{M}}^{1/f}\lp\JD_2\rp}
{Q_{w=x\sqrt{M}}^{1/f}\lp\JD_1 \rp}\right]^2 \; .
\end{equation}
A non-trivial check will be that the luminosity ratios obtained with
these two different expressions are consistent with each other. We
shall see below that this is generally the case.

\section{Results without pileup}
\label{sec:results}

\begin{figure}[p]
  \centering
  \includegraphics[width=0.8\textwidth]{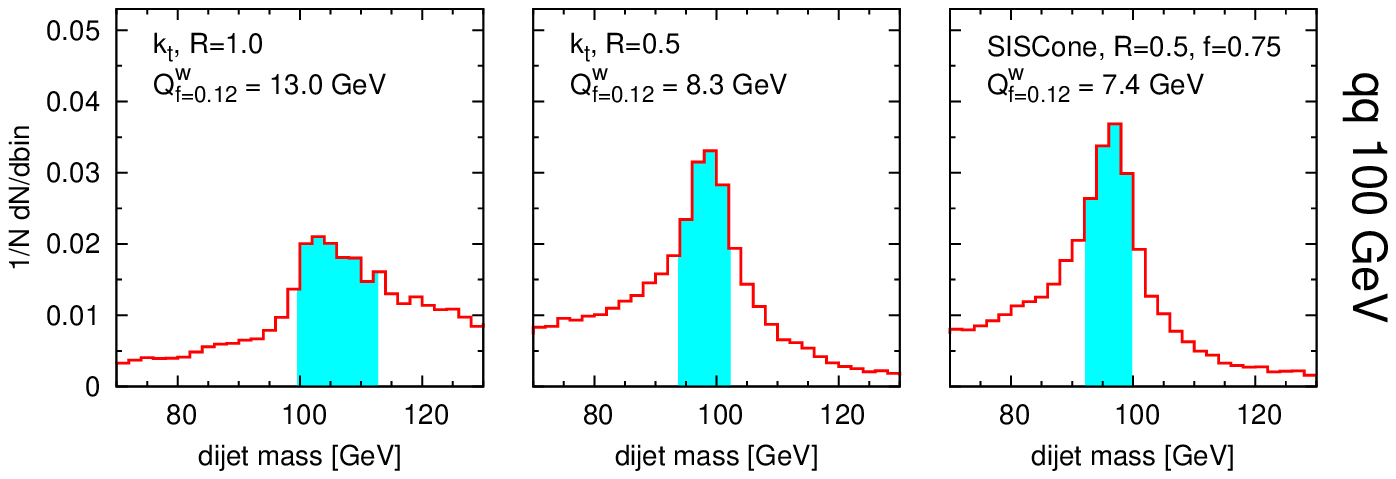}
  \includegraphics[width=0.8\textwidth]{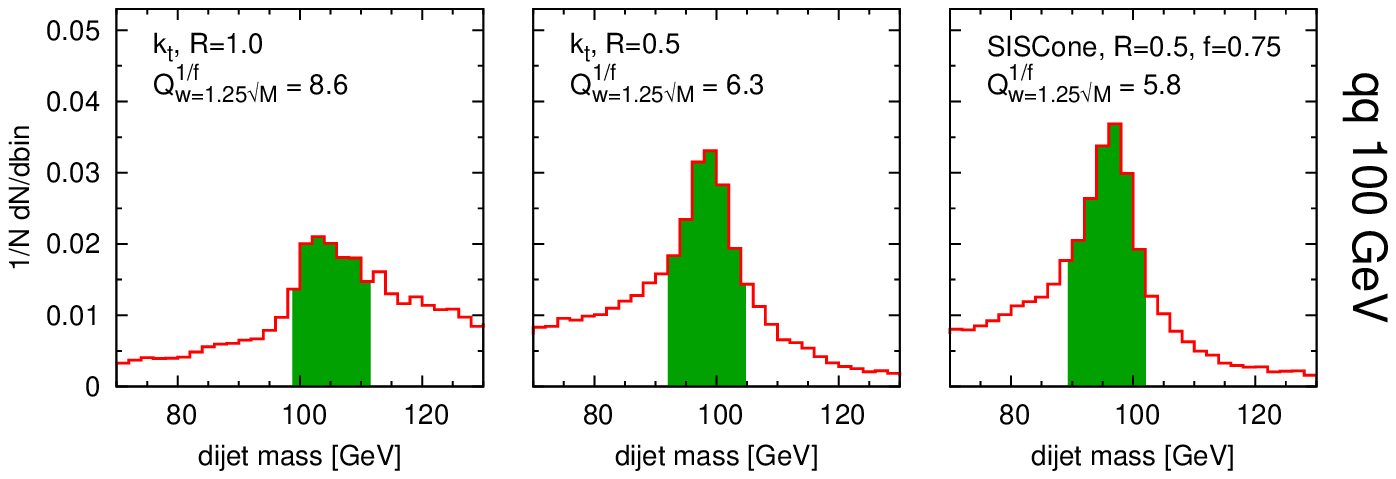}\vspace{1cm}

  \includegraphics[width=0.8\textwidth]{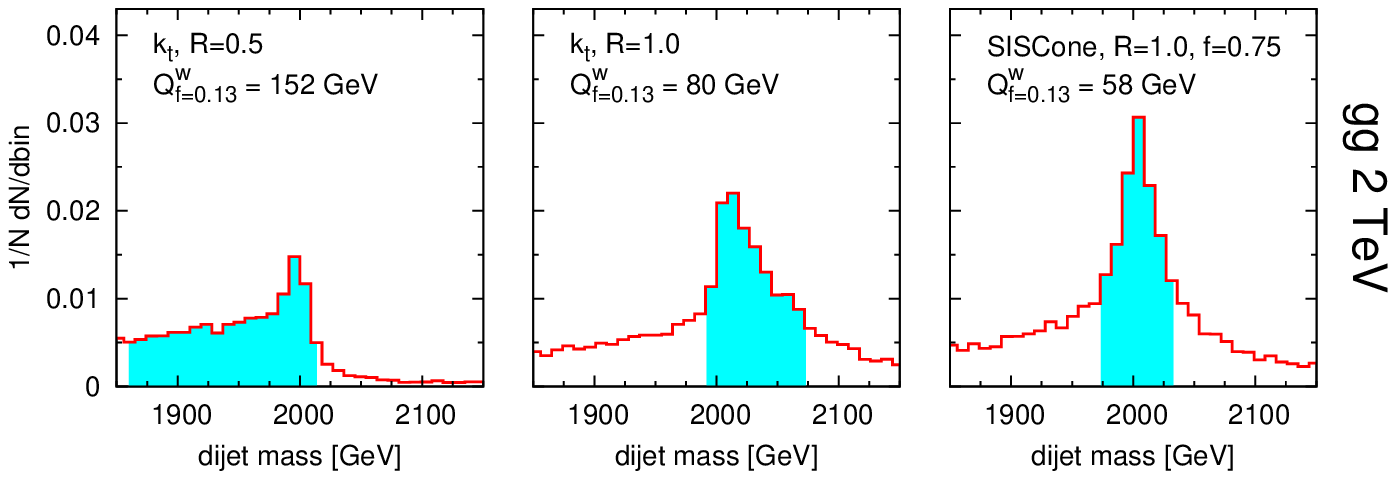}
  \includegraphics[width=0.8\textwidth]{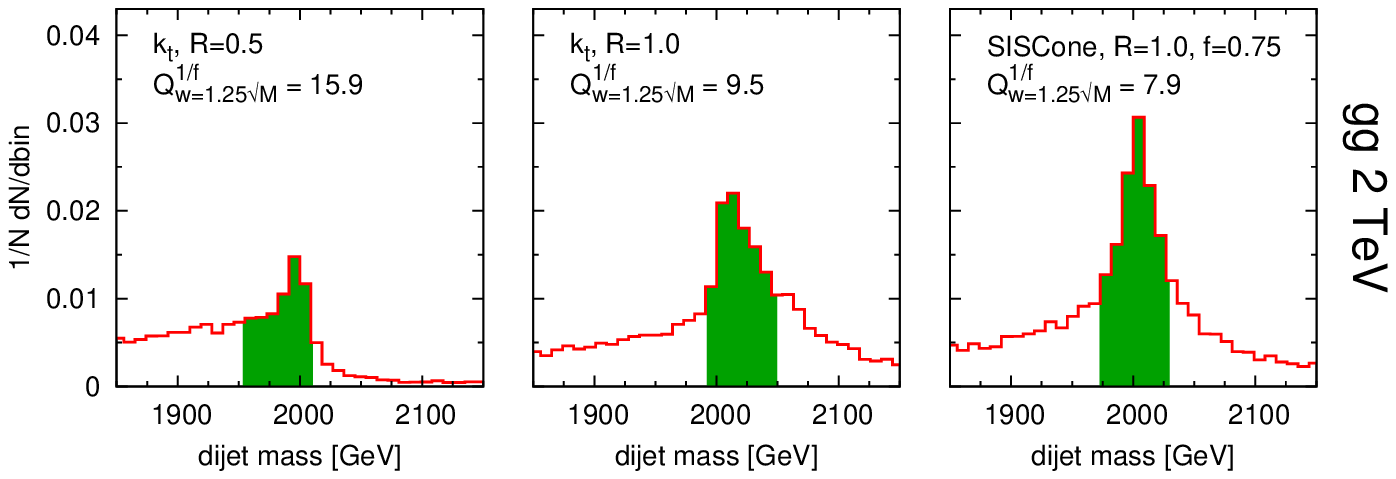}
  \caption{Illustrative dijet invariant mass distributions for two
    processes (above: \qq case at $M=100\GeV$; below: \gg case
    at $M=2\TeV$), comparing three jet definitions for each
    process. The shaded bands indicate the regions used when obtaining
    the two different quality measures. Note that different values of
    $R$ have been used for the \qq and \gg cases. }
  \label{fig:histos-no-PU}
\end{figure}

Let us start by illustrating the quality measures of
section~\ref{sec:figures_of_merit} for two examples of the
processes discussed in section~\ref{sec:ev-gen}.
Figure~\ref{fig:histos-no-PU} shows dijet invariant mass distributions
for the $100\GeV$ \qq case (upper 6 plots) and the $2\TeV$ \gg case
(lower 6 plots). In each case we show 3 different jet
definitions. Together with the histograms, we have included a shaded
band that represents the region used to calculate the quality
measures. 
In the first and third row we consider $Q_{f=z}^w$ and
the quality measure is given by the width of the (cyan) band.
In the second and fourth rows, the histograms are the same, but we now
show the (dark-green) band used in determining
$Q^{1/f}_{w=1.25\sqrt{M}}$ --- the quality measure is given by the
total number of generated events, divided by the number of events
contained in the band.

\begin{figure}[p]
  \centering
  \includegraphics[width=0.48\textwidth]{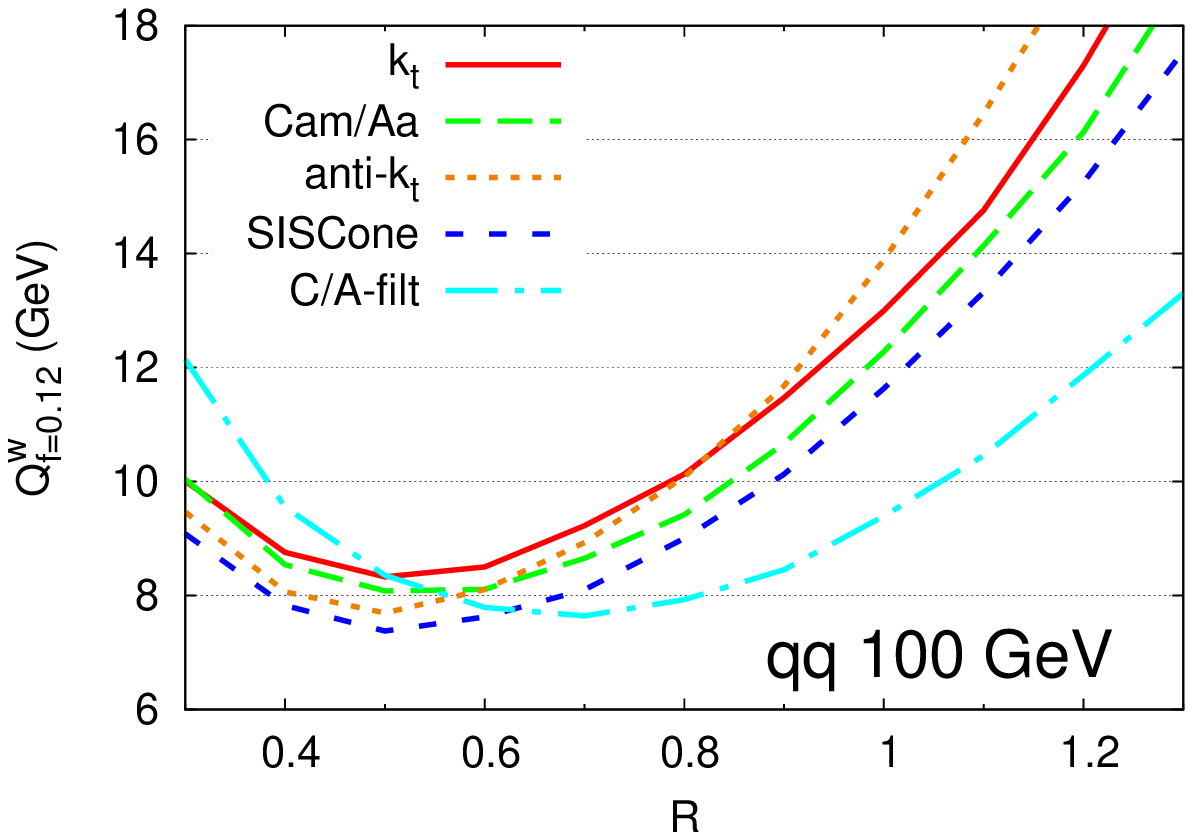}
  \includegraphics[width=0.48\textwidth]{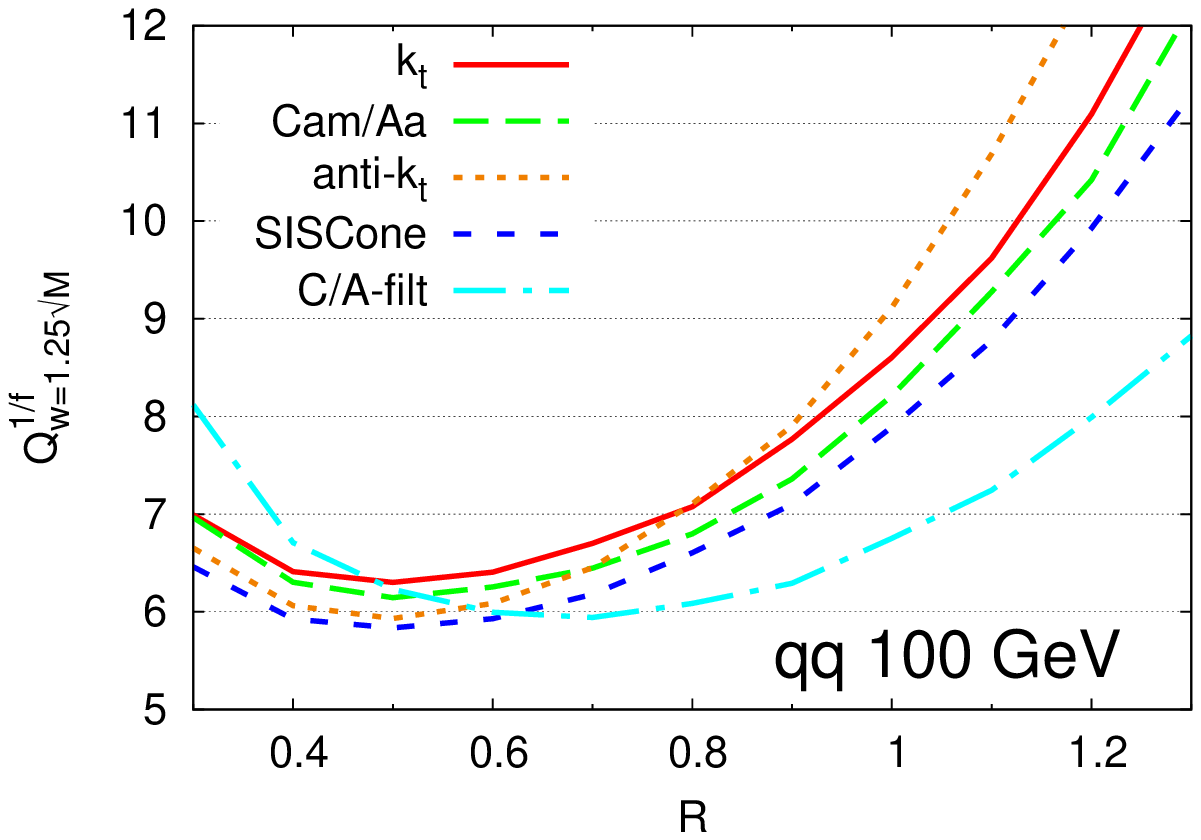}

  \includegraphics[width=0.48\textwidth]{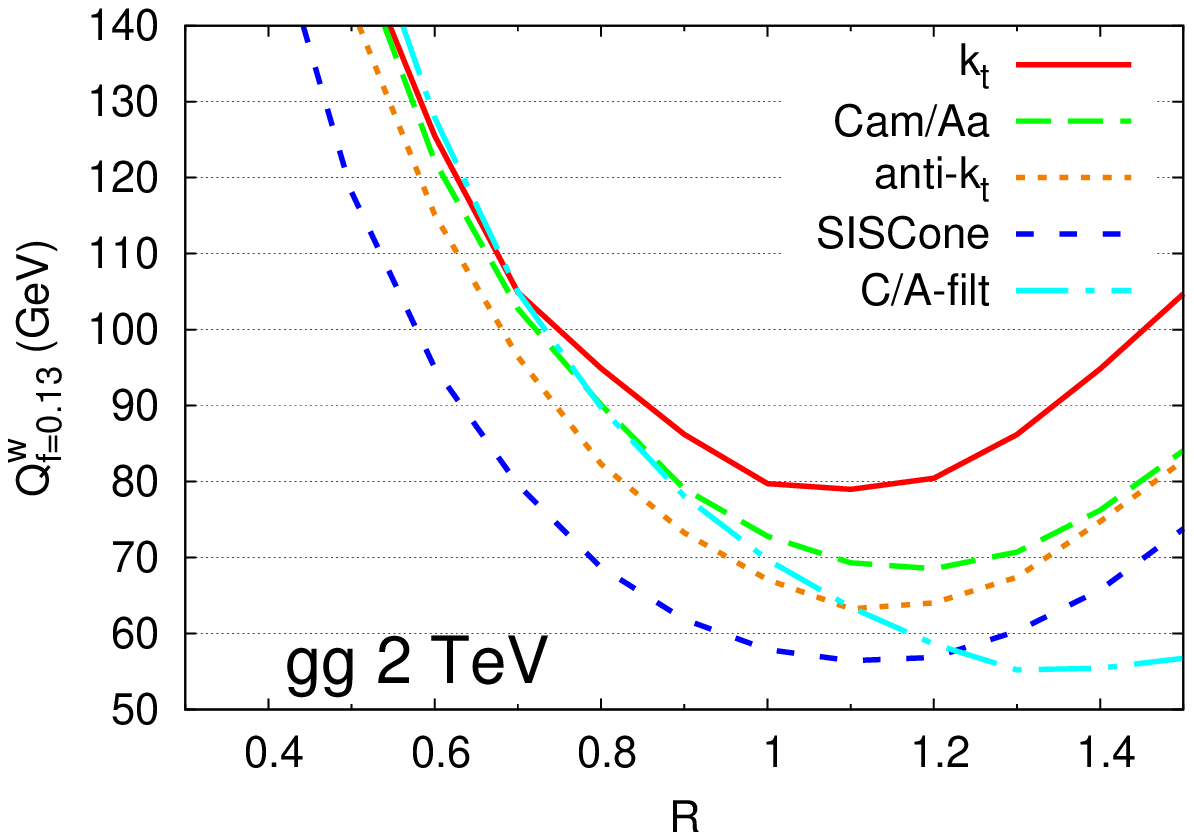}
  \includegraphics[width=0.48\textwidth]{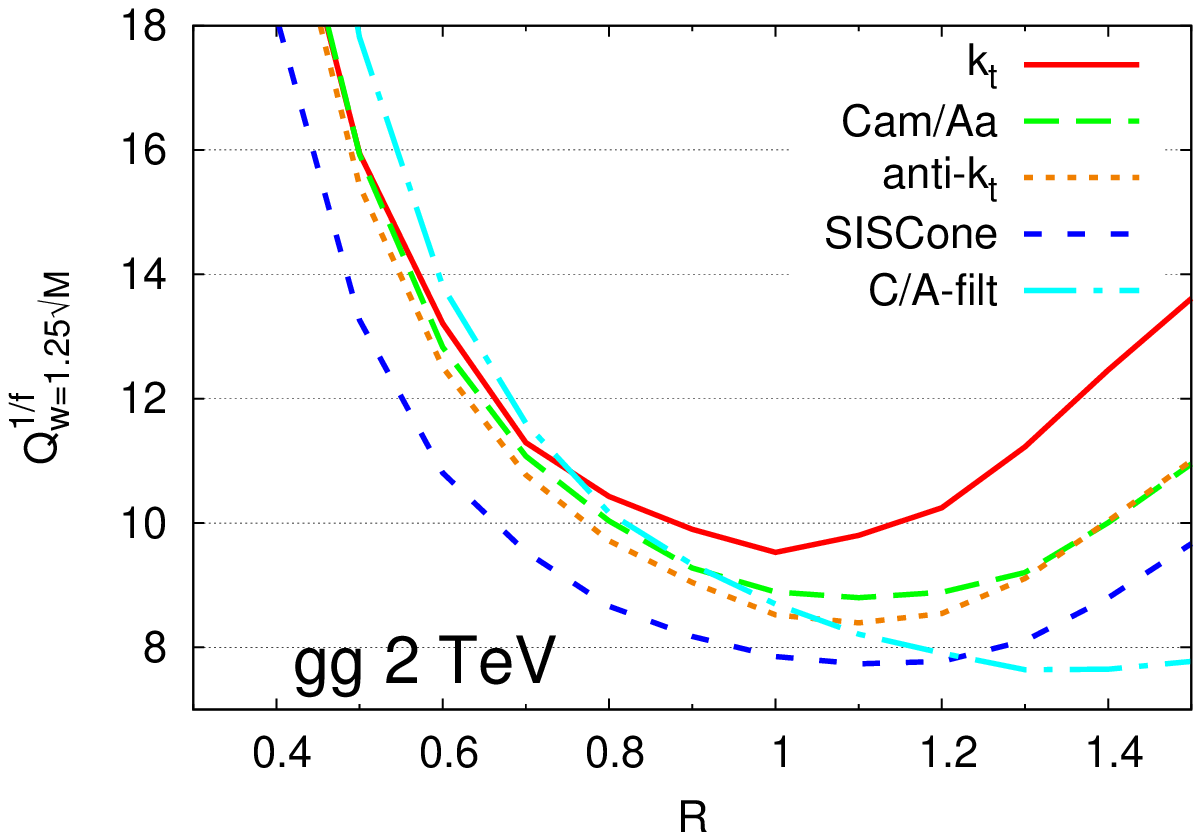}

  \includegraphics[width=0.48\textwidth]{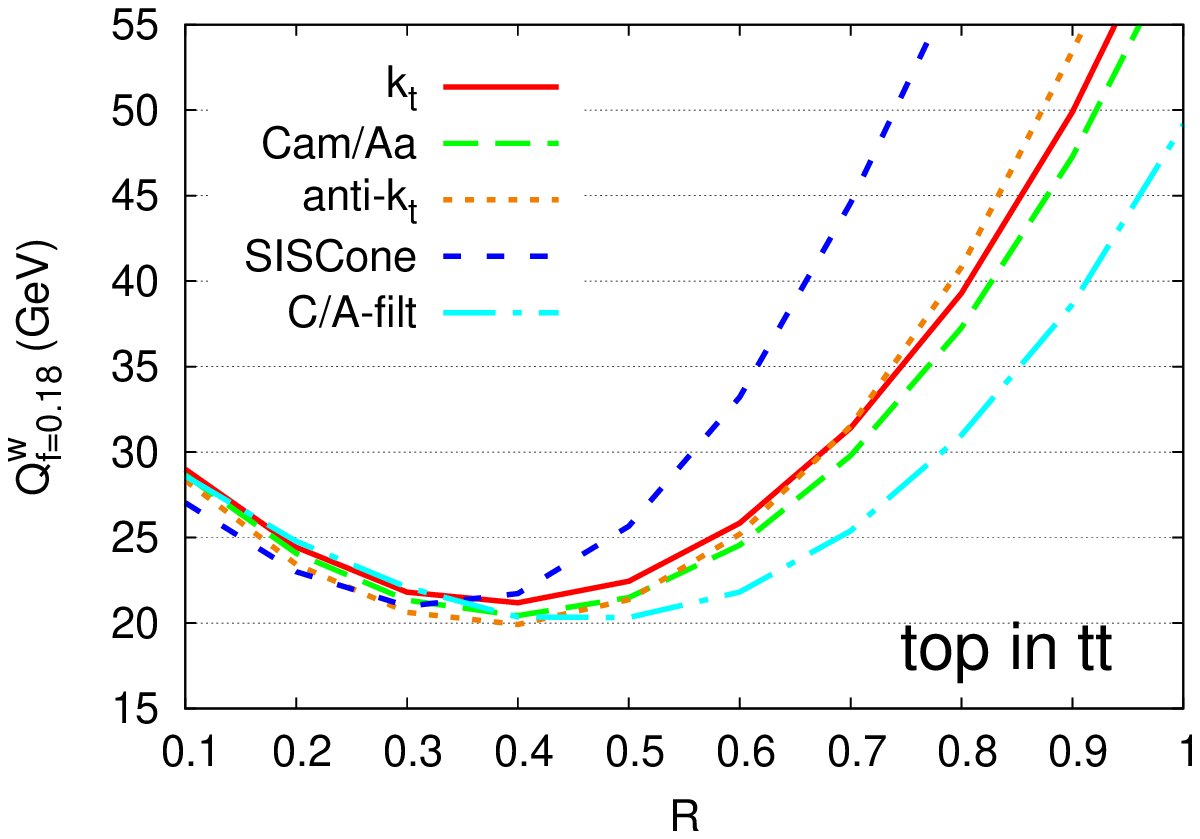}
  \includegraphics[width=0.48\textwidth]{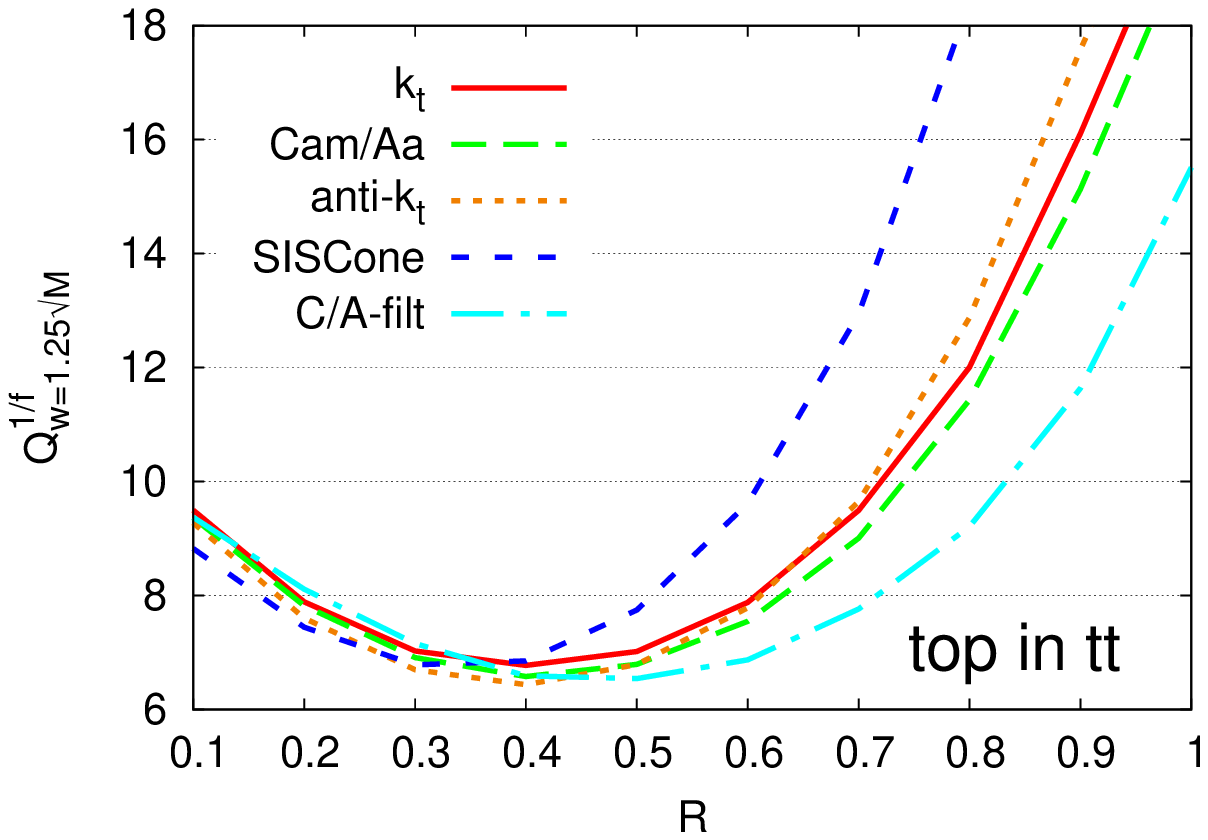}
  \caption{\small The quality measures \Qa{z} (left) and
    \Qb{1.25} (right), for different jet algorithms as a function of
    $R$, for the $100\GeV$ \qq case (top row),  $2\TeV$ \gg (middle
    row) and top reconstruction in $t\bar t$ events (bottom row).}
  \label{fig:all_merit} 
\end{figure}

Within a given
row of figure~\ref{fig:histos-no-PU} (same process, but different jet definitions), the histograms that
``look'' best (\ie the rightmost plots) are also those with the
smallest quality measures, as should be the case. Furthermore in the
situations where one histogram looks only moderately better than
another (\eg top row, central and right plots), the values of the
quality measures are appropriately close. This gives us a degree of
confidence that the quality measures devised in the previous section
behave sensibly, and provide a meaningful numerical handle on the
otherwise fuzzy concept of ``best-looking''.

Before moving on to a more systematic studies of how the quality
measures depend on the choice of jet definition, we observe that in
figure~\ref{fig:histos-no-PU}, smaller $R$ gives better results for
the $100\GeV$ \qq case, while for the the $2\TeV$ \gg case, the opposite
happens. This is a concrete illustration of the fact that there is no
universal best jet definition.

\begin{figure}[t]
  \centering
  \includegraphics[width=0.48\textwidth]{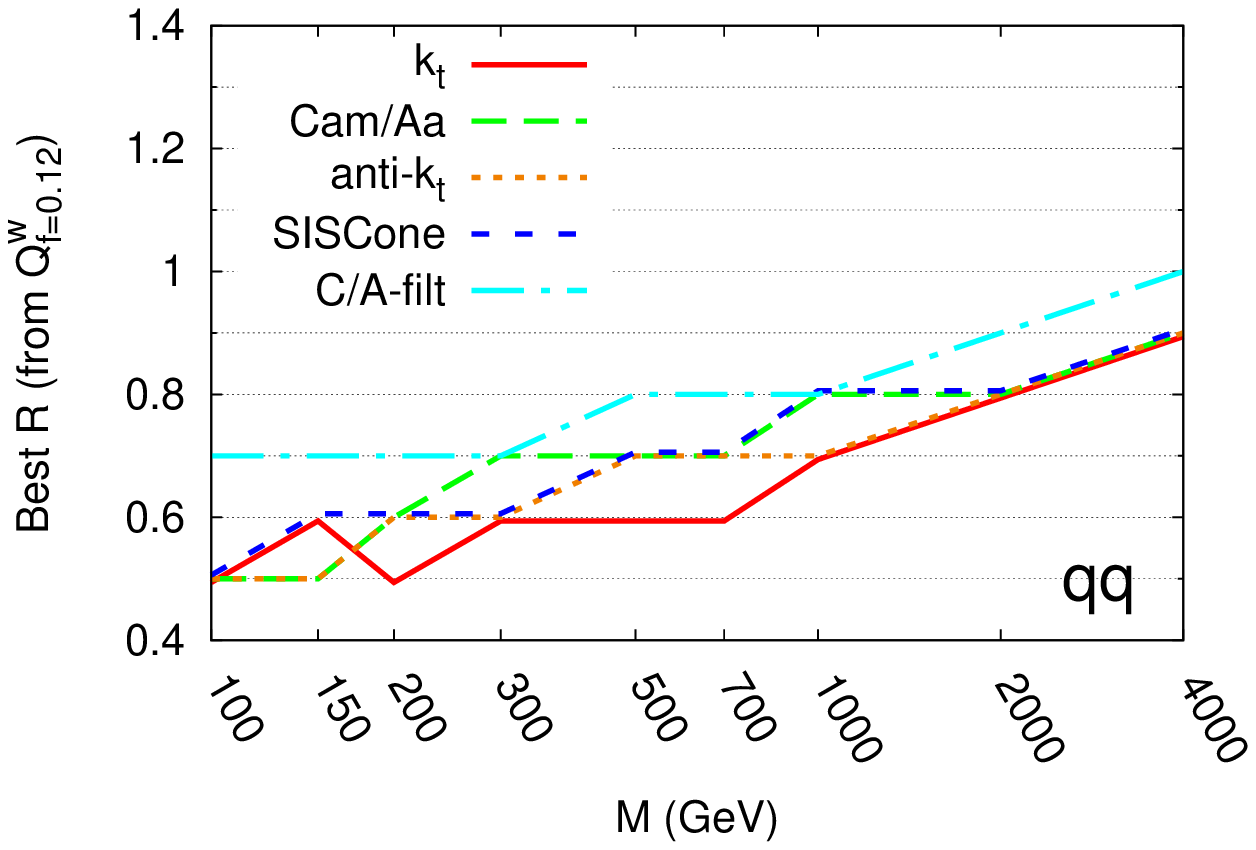}
  \includegraphics[width=0.48\textwidth]{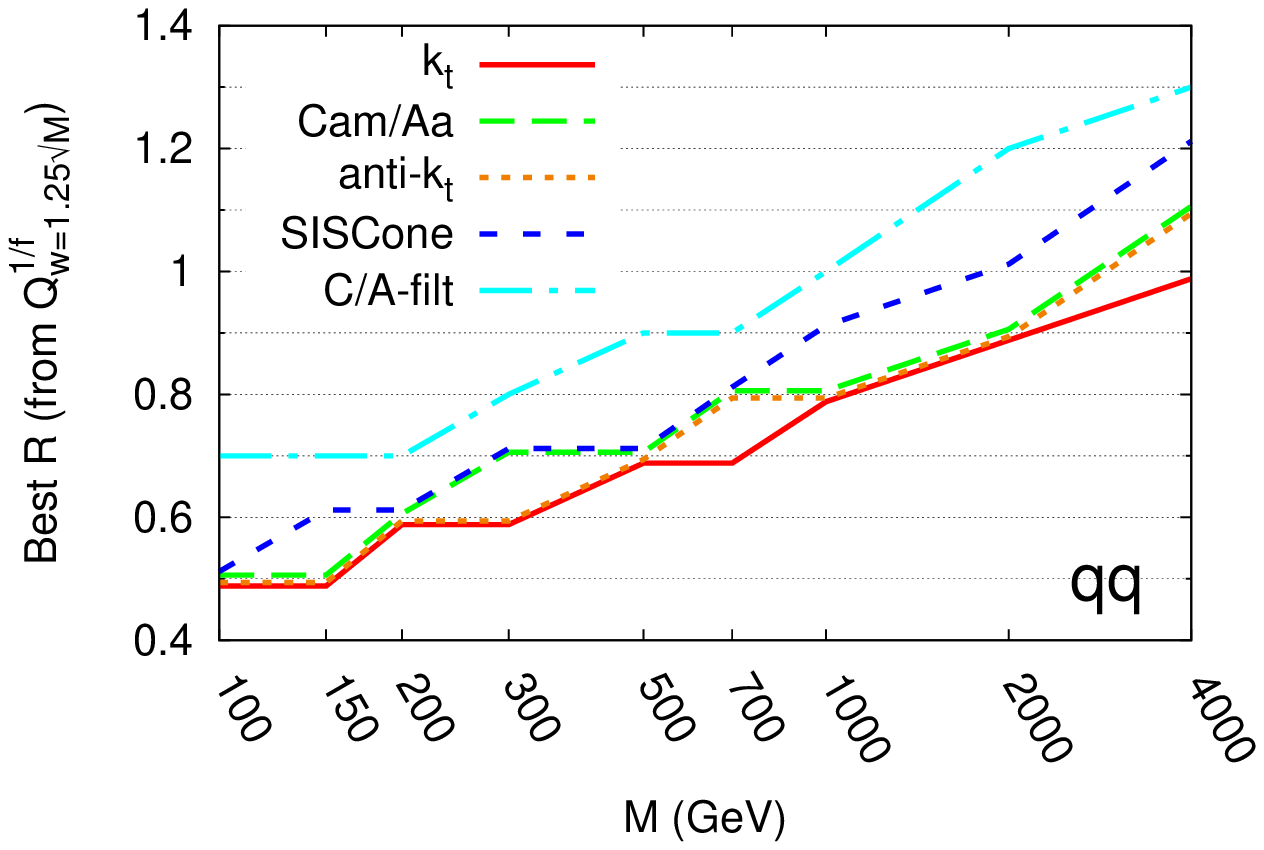}
  \includegraphics[width=0.48\textwidth]{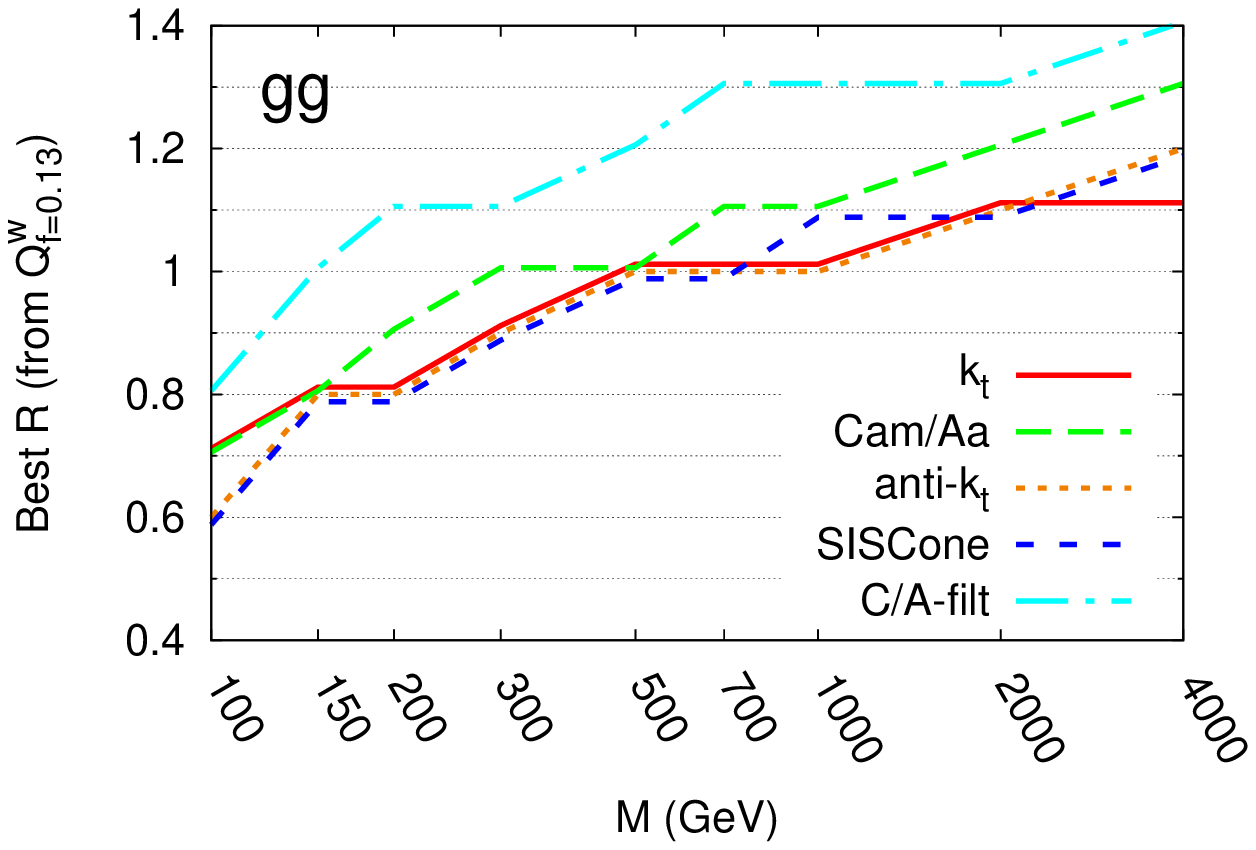}
  \includegraphics[width=0.48\textwidth]{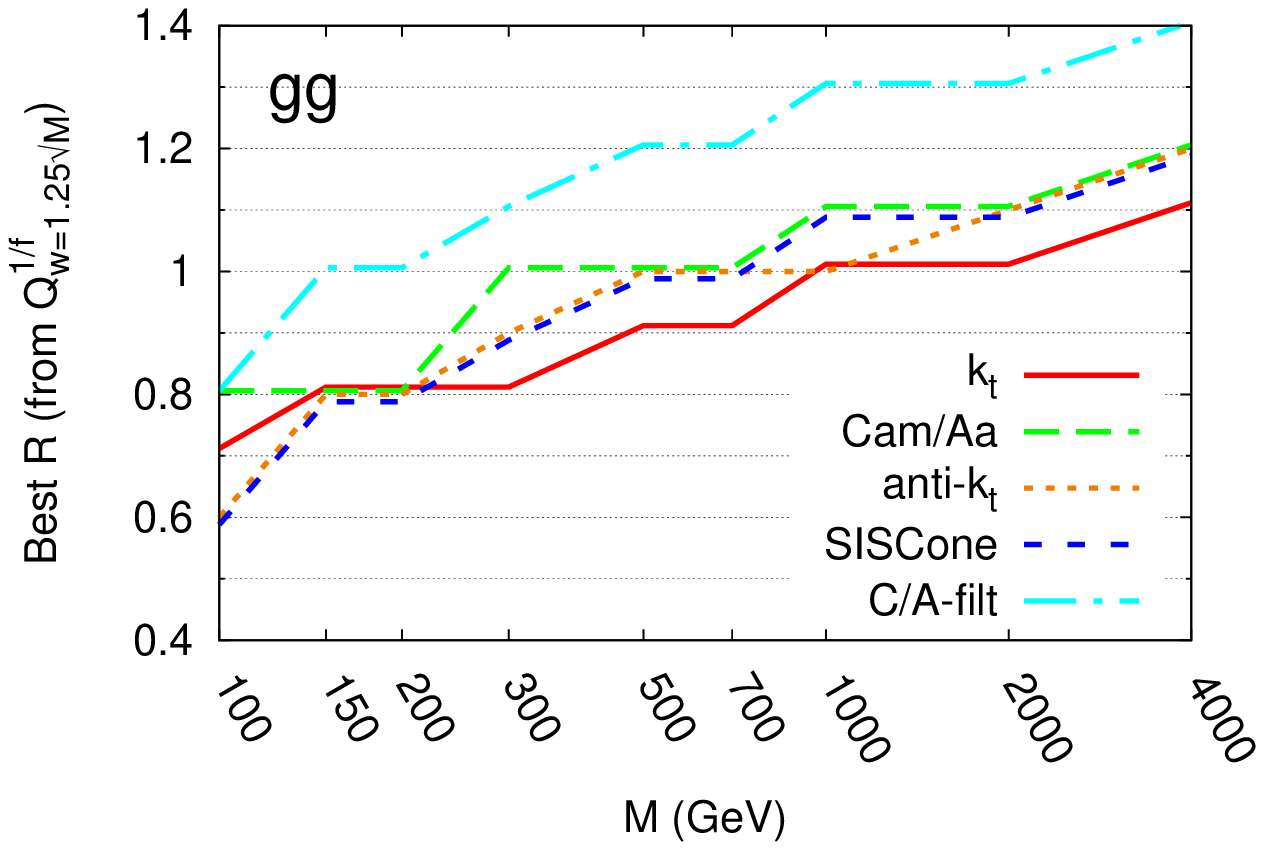}
  \caption{\small The optimal value for $R$ as a function of the mass
    of the \qq/\gg system  (upper/lower rows), as determined from the two
    quality measures  (left, right columns) for various jet algorithms. }
  \label{fig:Rbest-no-PU} 
\end{figure}

Next, in figure~\ref{fig:all_merit}, we show the values of the two
quality measures (left: \Qa{z}, right \Qb{1.25}) for different
jet algorithms as a function of $R$. The top and middle rows correspond
to the two processes already studied in figure~\ref{fig:histos-no-PU}
($100\GeV$ \qq and $2 \TeV$ \gg), while the bottom row corresponds to
top reconstruction in $t\bar t$ events. These plots allow one to
compare the different jet algorithms, and for each one to determine
the radius value that gives the best quality measure. The curves
confirm the earlier observation that the $2\TeV$ \gg case prefers a
substantially larger choice of $R$ than the $100\GeV$ \qq case. 
To understand this characteristic, it is useful to
consider figure~\ref{fig:Rbest-no-PU}, which gives the best $R$ (\ie
position of the minimum of the quality measure for each algorithm) as
a function of momentum scale, separately for the quark and gluon
cases.
There one sees that for gluonic jets one prefers a larger $R$ than for
quark jets, and one also prefers a larger $R$ as one moves to higher
momentum scales. 

This general pattern was predicted in~\cite{Dasgupta:2007wa}, and is
understood in terms of an interplay between the jet needing to capture
perturbative radiation, but without excessive contamination from
underlying-event (UE) ``noise'': whereas perturbative arguments alone
would favour $R$ of order $1$, the need to limit the amount of UE in
the jet pushes one to lower $R$. The UE matters most relative to
the jet energy for low-$p_t$ jets, and perturbative radiation matters
more for gluon jets.

A further remark is that the optimal values of $R$ found for processes
involving $\sim100 \GeV$ mass scales, $R \sim 0.5$, correspond quite
closely to values used typically by the Tevatron experiments and in
many LHC studies (see \eg\ \cite{Ellis:2007ib,Bhatti:2008hz}).\footnote{For the Tevatron there will actually be a
  preference for slightly larger $R$ values than at LHC, a consequence
  of the more modest UE.}
Our analysis here confirms that those are therefore
good choices. However, at the high scales that will be probed by LHC,
$\sim 1\TeV$, our results indicate that it is important for the
experiments to use jet definitions with somewhat larger values of
$R$.

\begin{figure}[p]
  \centering
  \includegraphics[width=0.8\textwidth]{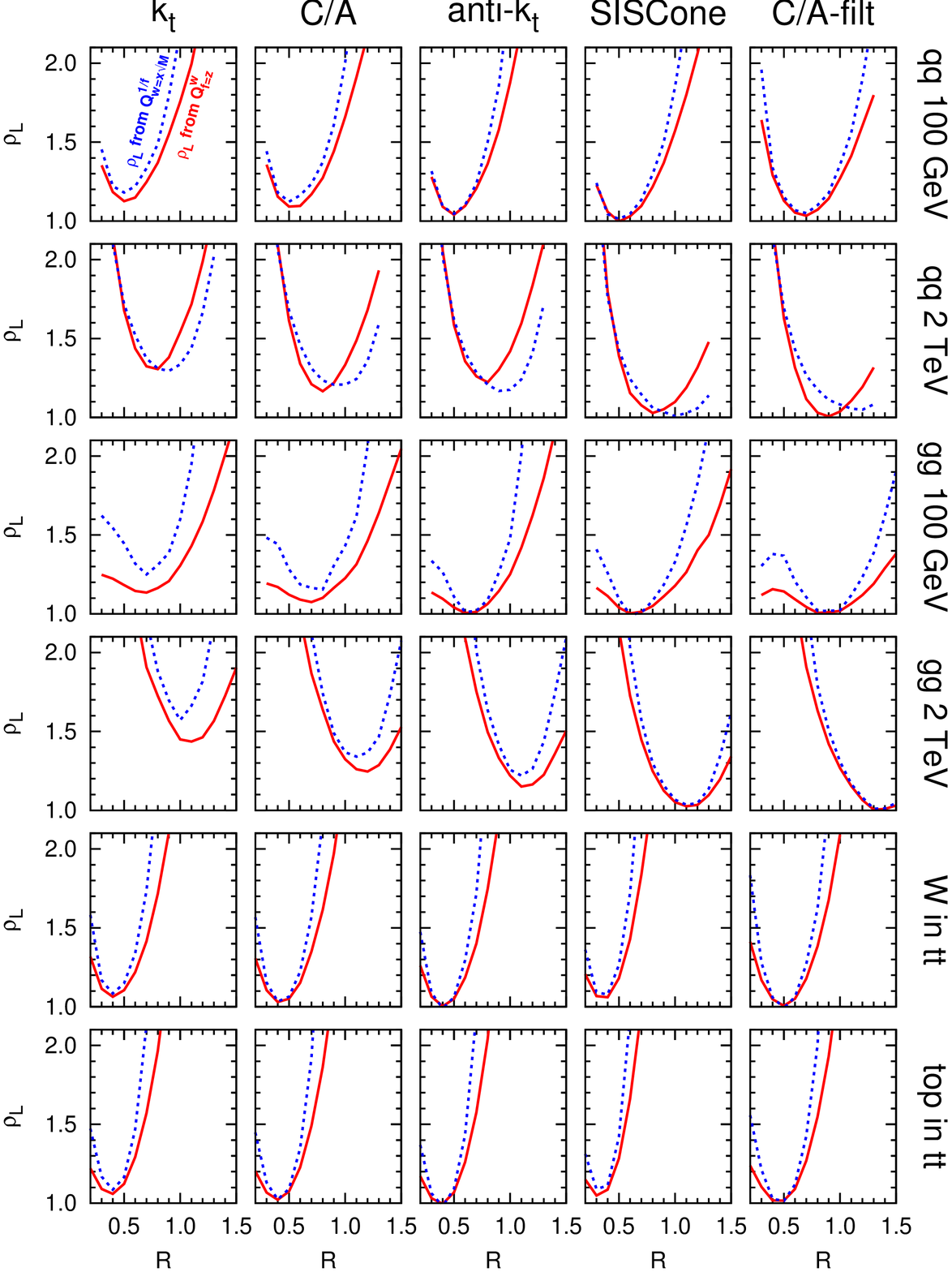}
  \caption{
    For each process (one per row)
    this plot shows the luminosity ratio $\rhoL$ required in order to
    obtain the same significance as with the best jet definition. The (red)
    solid line corresponds to the estimate of $\rhoL$ from
    eq.~(\ref{eq:rhol_minwidth}) (based on the minimal width \Qa{z}),
    while the (blue) dotted line corresponds to
    eq.~(\ref{eq:rhol_maxfrac}) (based on the maximal fraction
    \Qb{1.25}). }
  \label{fig:summary-no-PU}
\end{figure}

The quantitative impact of a poor choice of jet definition is
illustrated in figure~\ref{fig:summary-no-PU}. For each process, we have
identified the jet definition, $\JD_{\rm best}$, that provides the
best (lowest) value of the quality measure (\cnf
table~\ref{tab:best}). Then for every other jet 
definition, $\JD$, we have calculated the effective
increase in luminosity, $\rhoL(\JD/\JD_{\rm best})$ as in
eq.~(\ref{eq:rhol_basic_def}), that is needed to obtain as good a
significance as with JD$_{\rm best}$. This is shown for each jet
algorithm as a function of $R$, with (red) solid lines for \Qa{z},
using eq.~(\ref{eq:rhol_minwidth}), and (blue) dashed lines for
\Qb{1.25}, using eq.~(\ref{eq:rhol_maxfrac}).

\begin{table}
\begin{center}
\begin{tabular}{|l|cl|c|cl|c|}
\hline
& \multicolumn{3}{c|}{\Qa{z}} & \multicolumn{3}{c|}{\Qb{1.25}}\\
\cline{2-7}
Process          & \multicolumn{2}{c|}{$\JD_{\rm best}$} & $Q$ (GeV)   &
\multicolumn{2}{c|}{$\JD_{\rm best}$} & $Q$ \\
\hline
\qq, 100 GeV     & SISCone& $R$=0.5    & 7.38  & SISCone& $R$=0.5    & 5.83  \\
\qq, 2 TeV       & C/A--filt& $R$=0.9  & 20.8  & SISCone& $R$=1.0    & 5.18    \\
\gg, 100 GeV     & SISCone& $R$=0.6    & 14.7  & SISCone& $R$=0.6    & 8.78  \\
\gg, 2 TeV       & C/A--filt& $R$=1.3  & 55.2  & C/A--filt& $R$=1.3  & 7.64   \\
$W$ in $t\bar t$ & anti-$k_t$& $R$=0.4 & 10.7  & anti-$k_t$& $R$=0.4 & 5.37  \\
$t$ in $t\bar t$ & anti-$k_t$& $R$=0.4 & 19.9  & anti-$k_t$& $R$=0.4 & 6.44  \\
\hline
\end{tabular}
\end{center}
\caption{The $\JD_{\rm best}$ jet definitions for the various
  processes of figure~\ref{fig:summary-no-PU}, together with the
  corresponding  $Q(\JD_{\rm best})$ values used in calculating
  $\rhoL$. In the $2\TeV$ \qq case, the $\JD_{\rm best}$ definitions
  differ, but figure~\ref{fig:summary-no-PU} shows that they lead to
  very similar quality measures, and the
  question of
  which is ``best'' ultimately depends on fine details of their behaviour.
}
\label{tab:best}
\end{table}

A first observation is that in general, the two quality measures lead
to similar results for $\rhoL$. This is a non-trivial check that the
procedure is consistent and that our quality measures behave
sensibly.\footnote{In a few instances there are moderate differences
  between the two determinations of $\rhoL$.
  This usually occurs when the resulting window widths for the two
  measures differ substantially (\ie they probe the distribution with
  different effective resolutions).
  These cases, however, do not significantly alter any conclusions.}
One should be aware that there is some degree of arbitrariness in the
choice of $z$ for \Qa{z} and $x$ for \Qb{x}. Accordingly, we
have also examined results for the case where these choices are
doubled, and verified that $\rhoL$ is again similar (in this respect
\Qb{x} seems to be somewhat more stable, \cnf the web-pages at
\cite{quality.fastjet.fr}).

Next, let us discuss the impact of using the worst jet algorithm (at
its best $R$) compared to the best jet definition. At small energy
scales one requires about $10-20\%$ extra luminosity, a modest effect.
At high masses this increases to $30-40\%$. In general it seems that
SISCone and C/A-filt are the best algorithms (and are similar to each other), while the
$k_t$ algorithm fares worst. In some cases anti-$k_t$ also performs
optimally.

The penalty for choosing a non-optimal $R$ can be even larger. For
example, using SISCone with $R=0.4$ ($0.5$) at $2\TeV$ leads to
$\rhoL$ of about $1.75$ ($1.35$) for the \qq case, and $\sim 3$ ($2$)
for the \gg case.
The use of $R\simeq 0.4-0.5$ is widespread in current LHC analyses
(for example, \cite{Bhatti:2008hz} used $R=0.5$ with the CMS iterative
cone,
which is similar to anti-$k_t$) and if this is maintained up to high
mass scales, it may  lead to a need for twice as much integrated
luminosity (or even more) to make a discovery as with an optimised
choice of jet definition.

A point worth bearing in mind is that the quality measures do not
provide all relevant information about the peak. For example, for the
small-mass gluonic case, the smallest $R$ values do not lead to
appreciably worse-than-optimal $\rhoL$ results, however if one
examines the position of the peak (\cnf the histograms available via
the web-tool
\cite{quality.fastjet.fr}) one sees that it becomes rather unstable at
small $R$%
\footnote{This instability is the cause also of the (somewhat
  spurious) decrease of $\rho_L$ at small $R$ in the $100\GeV$ \gg
  C/A-filt case: there, the peak is at quite low masses ($\sim
  50\GeV$), and the limited phase-space towards yet-smaller masses
  causes it to narrow slightly as one further reduces $R$.
  Cases like this are rare and  easy to identify
  when they occur.  } %
\cite{ORprivate}.
Nevertheless, in most cases, and in particular for all but a few
pathological regions of the jet-definition parameter-space, the
position of the peak is sensible.

So far we have discussed only simple, dijet events. The last row of
figure~\ref{fig:all_merit} and the last two rows of
figure~\ref{fig:summary-no-PU} show results for more complex $t\bar t$
events, which here decay to 6 jets.
A first observation is that the optimal $R$ is fairly similar to that
in the 100$\GeV$ \qq case --- this is perhaps not surprising, since in
both cases the energy of each jet is around 50$\GeV$.
More detailed inspection shows however that the range of
``acceptable'' $R$ is significantly smaller for the $t\bar t$ case
than the \qq case. This is most visible in
figure~\ref{fig:summary-no-PU}, and especially for the SISCone
algorithm. The reason is simple: in multijet events, as $R$ is
increased, jets that should represent distinct leading-order
``partons'' may end up being merged.\footnote{SISCone's higher
  sensitivity to this effect is a consequence of the fact that for a
  given $R$ value it can cluster two hard particles that are up to
  $2R$ apart, whereas the other algorithms reach out only to $R$. }
 
This issue should be kept in mind when planning analyses involving
multijet final states at high energy scales --- with current
algorithms, there will then be a significant tension between the need
for a large radius owing the high energy scale, and the need for a
small radius in order to disentangle the many jets. 
In this respect, methods that use parameters other than just $R$ to
resolve jet structure (including some originally intended for jet
substructure), such as
\cite{Butterworth:2002tt,Seymour:2006vv,Butterworth:2007ke,Butterworth:2008iy,Thaler:2008ju,Kaplan:2008ie,Almeida:2008yp},
are likely to be of prime importance in obtaining optimal jet results.
The detailed general investigation of such ``third-generation''
jet-methods\footnote{\label{foot:generations} We use the term ``first
  generation'' jet-methods for the infrared and/or collinear unsafe
  cone algorithms of the '80s and '90s, ``second generation'' for
  subsequent infrared and collinear safe methods 
  (recombination algorithms like JADE, $k_t$, C/A and anti-$k_t$, as
  well as the cone algorithm SISCone) 
  that essentially use one main fixed parameter to
  specify the resolution for jets. By third generation methods, we
  have in mind those that exploit more than a single ``view'' of the
  event, and which may be based on a more powerful use of
  existing algorithms.} %
deserves further work.

A final comment concerns studies to reconstruct boosted top quarks,
for example from high-mass resonances that decay to $t\bar t$ (see \eg
\cite{Agashe:2006hk,Lillie:2007yh,Baur:2008uv}). Significant recent
work has gone into investigating the identification of top quarks in
this context
\cite{BroojimansTop,Thaler:2008ju,Kaplan:2008ie,Almeida:2008yp,Almeida:2008tp},
where their decay products are often contained within a single jet. In
such a situation, the best $R$-value for carrying out a top-ID subjet
analysis depends on the top $p_t$, as in \cite{Kaplan:2008ie} (similar
to the C/A-based subjet Higgs search of \cite{Butterworth:2008iy}),
and is conditioned by the need to take a jet opening angle
commensurate with the top-quark dead-cone size and decay angle,
$\order{2m_t/p_t}$ ($p_t \sim M/2$ where $M$ is the resonance mass), so that
the jet contains the top decay products, but not gluon emission from
the top quark itself (which would smear and skew the top mass
reconstruction). However, to obtain a good mass-resolution on the
high-mass resonance that is reconstructed
from the $t\bar t$ system, it is instead necessary to \emph{include}
any gluon radiation from the top quark inside the jets, and for this
purpose, optimal $R$ values will be the larger ones found here for
normal $q\bar q$ dijet-events and will grow with $p_t$ (this issue
obviously does not arise for boosted electroweak bosons).
Thus, and in contrast to what has been investigated so far, in the work
referred to above, one should work simultaneously with two $R$ values:
a small one, $\order{2m_t/p_t}$ for identifying the top-quark decays, and
a larger one, as given by figure~\ref{fig:Rbest-no-PU}, for
determining the top-quark momentum just as it was produced from the
resonance decay.
In this respect C/A-based solutions (including
filtering) are particularly interesting insofar as they allow
consistent views of an event at multiple $R$-values.

\section{Results with pileup}
\label{sec:results-pu}

It is foreseen that the LHC will operate at a range of different
luminosities. One should therefore establish whether the
conclusions of the previous section are robust in the presence of
multiple minimum-bias (MB) pileup (PU) events.
We will consider low and high luminosity scenarios: $\mathcal{L}_{\rm
  low}=0.05$ mb$^{-1}$ and $\mathcal{L}_{\rm high}=0.25$ mb$^{-1}$ per
bunch crossing,\footnote{Even larger luminosities, $\mathcal{L}_{\rm
    vhigh}=2.5$ mb$^{-1}$, might be relevant at the sLHC upgrade (see
  for example \cite{sLHC}).} corresponding respectively to an average
of about 5 and 25 minimum-bias collisions per bunch crossing (or
instantaneous luminosities of
$2\times10^{33}\unit{cm}^{-2}\unit{s}^{-1}$ and 
$10^{34}\unit{cm}^{-2}\unit{s}^{-1}$
with a $25\,\mathrm{ns}$ bunch 
spacing). The MB events are simulated using {\tt Pythia} 6.410
\cite{Sjostrand:2006za} with the DWT tune~\cite{Albrow:2006rt}, as for
our hard event samples, and the number of MB events added to a specific
hard event has a Poisson distribution.

It is well known that PU degrades mass resolution and shifts the
energy scale, and the LHC experiments will attempt to correct for
this. Currently their procedures tend to be highly detector-specific,
which limits their applicability in a generic study such as ours. We will
therefore use the jet-area-based pileup subtraction method of
\cite{Cacciari:2007fd,Cacciari:2008gn}, which is
experiment-independent and straightforward to use within the {\tt FastJet}
framework.\footnote{Should the experiments' internal methods prove to
  be superior to the jet-area-based method, then the conclusions of
  this section will only be made more robust; if on the other hand
  they turn out to perform less well, there would be a compelling
  reason for them to adopt the area-based method.}
For completeness we give below a quick review of area-based
subtraction, and then we will examine the impact of pileup both with
and without its subtraction.

\subsection{Area-based pileup subtraction}

Area-based subtraction involves two elements: the calculation of jet
areas, which represent a given jet's susceptibility to contamination
from uniform background noise, and an estimate of the level of
background noise in the event.

As proposed in \cite{Cacciari:2008gn}, for each jet $j$ in an event,
one can determine a 4-vector area in the rapidity-azimuth plane
$A_{\mu j}$. Given an estimate for the amount $\rho$ of
transverse-momentum per unit area due to background noise, a jet's
corrected (\ie subtracted) momentum $p_{\mu j}^{(\rm sub)}$ is obtained as \cite{Cacciari:2007fd}
\begin{equation}
  \label{eq:pt-correct-4vect}
  \smash{p_{\mu j}^{(\rm sub)}} = p_{\mu j} - A_{\mu j}\, \rho \,.
\end{equation}
Subtracted jets are then used both for event selection and for the
mass reconstructions.

The quantity $\rho$ is taken to be independent of rapidity (an
acceptable approximation in much of the detector), and calculated on
an event-by-event basis as
\begin{equation}
  \label{eq:rho}
  \rho = \text{median} \left\{ \frac{p_{tj}}{A_{tj}}\right\}
  \,,
\end{equation}
where the median is obtained over all jets with $|y| <
5$.\footnote{The details of the peak position after subtraction (not
  the main subject of our study here) can depend on the choice of jets
  used for calculating $\rho$, with the issue being largest for large
  $R$.
  As an example, in the $2\TeV$ \gg case with $R=1$, the residual
  impact on the peak position can be of order $10\GeV$.
}
Regardless of the jet-definition used to analyse
the hard event, $\rho$ is always calculated using jets obtained with
the $k_t$ algorithm with $R=0.5$, which choice was found to be
particularly robust in~\cite{Cacciari:2007fd}.\footnote{Other
  technical settings are as follows: we use the active area for all
  jet algorithms except SISCone, which for speed reasons we use with the
  passive area. The area of ghost particles is set to $0.01$, they
  cover the range $|y|<6$, and the repeat parameter is 1.
  All other parameters correspond to the defaults in
  {\tt FastJet} 2.3. For C/A with filtering, the subtraction is carried out
  before the filtering stage. } %

\subsection{Results}

\begin{figure}[p]
  \centering
  \includegraphics[width=0.85\textwidth]{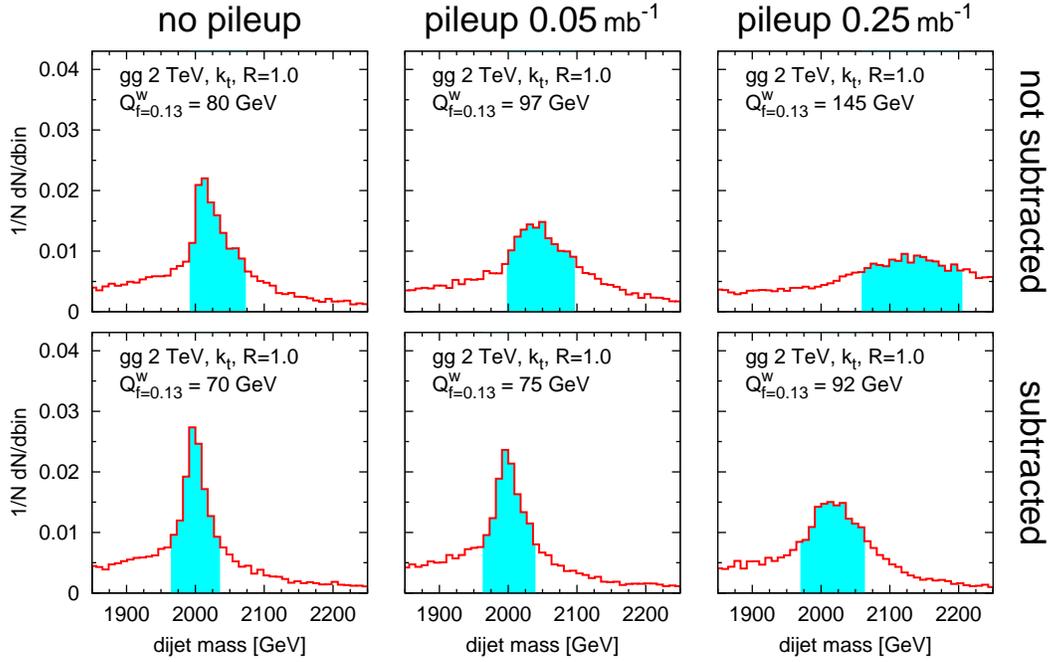}
  \caption{Invariant mass distributions for the $2\TeV$ \gg process,
    for the $k_t$ algorithm with $R=1$, shown with no pileup (left),
    low pileup (middle) and high pileup (right), without subtraction
    (upper row) and with subtraction (lower row). The shaded bands
    indicate the region used to calculate the \Qa{z} quality
    measure in each case.}
  \label{fig:histos-with-PU}
\end{figure}

Jets are most strongly affected by pileup at large $R$ values.
Accordingly, in figure~\ref{fig:histos-with-PU} we show histograms for
the $2\TeV$ \gg process, which without pileup favoured $R \gtrsim 1$.
The upper row shows results with no pileup, low and high pileup, all
without subtraction. One sees the clear degradation in the quality of
the peak as pileup is added, and this is reflected in the increasing
values of the quality measure. There is additionally a shift of the
peak to higher masses.

The lower row of figure~\ref{fig:histos-with-PU} shows the corresponding
results with subtraction. Note that we have applied subtraction even
in the case without pileup and one observes a non-negligible
improvement in the peak quality. This is because of the contribution
to the ``noisiness'' of the event from ``underlying-event'' (UE)
activity, which is in part removed by the subtraction procedure.
One sees even more significant improvements in the cases with pileup,
highlighting the importance of the use of some form of pileup subtraction.

\begin{figure}[t]
  \centering
  \includegraphics[width=0.8\textwidth]{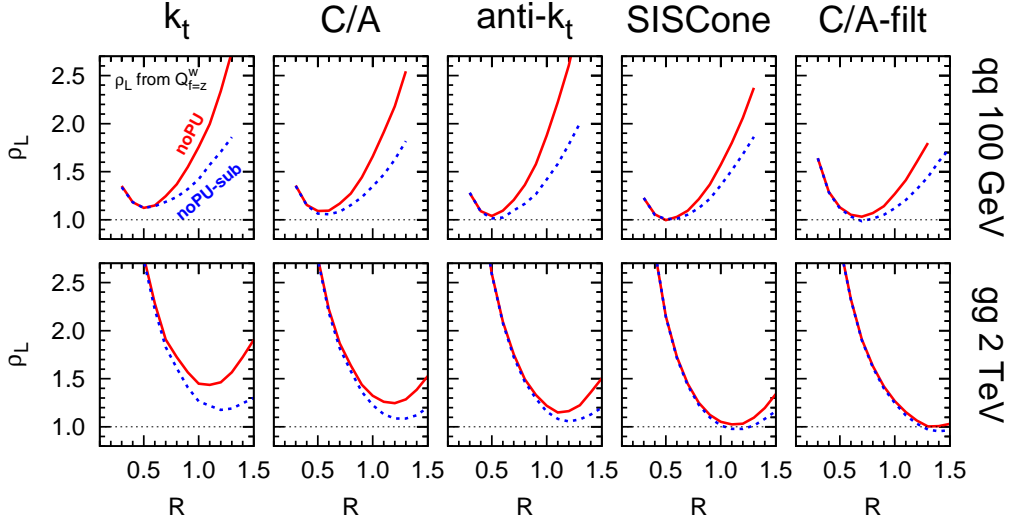}
  \caption{Illustration of the impact of subtraction in the absence of
    pileup. The effective luminosity ratios are based on the
    \Qa{z} measure, and normalised to the result for the best jet
    definition without pileup (no subtraction). The (red) solid curves
    show results without subtraction, the (blue) dotted curves with
    subtraction.}
  \label{fig:stamp-noPUsub-reduced}
\end{figure}

Let us now consider this more systematically. First, in
figure~\ref{fig:stamp-noPUsub-reduced} we examine the impact of
subtraction without pileup for all algorithms for the $100\GeV$ \qq and
$2\TeV$ \gg processes. This is given in terms of the effective
luminosity ratios normalised as in figure~\ref{fig:summary-no-PU}, \ie
to the lowest (best) value of the \Qa{z} quality measure across
all jet definitions without pileup (for brevity we omit results based
on \Qb{1.25}, which do not change the overall conclusions).
The (red) solid curve is always the same as in
figure~\ref{fig:summary-no-PU} (\ie for the jet algorithms without
pileup or subtraction), while the (blue) dotted curve shows the results with
subtraction. As expected, subtraction only matters significantly at
large $R$. The algorithms that performed worst without subtraction are
those that benefit the most from it, leading to quite similar optimal
quality for all algorithms.

\begin{figure}[p]
  \centering
  \includegraphics[width=0.8\textwidth]{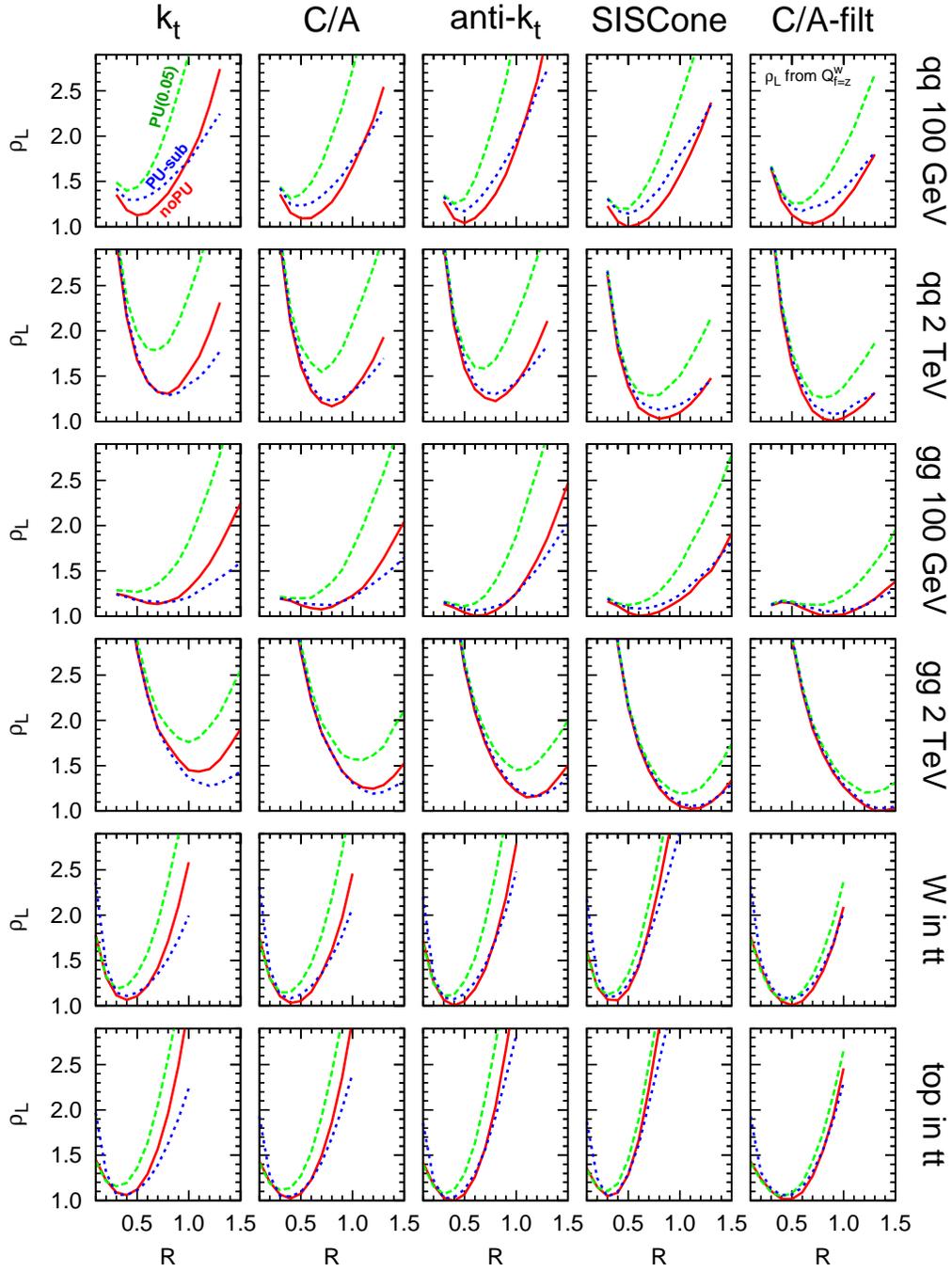}
  \caption{Illustration of the impact of low pileup, $0.05\mb^{-1}$
    per bunch crossing. Luminosity ratios have been calculated based
    on the \Qa{z} measure, and normalised to the result for the
    best jet definition without pileup (no subtraction). The (red)
    solid curves show the result with no pileup and no subtraction, the
    (green) dashed curves have pileup without subtraction and the (blue)
    dotted curves have pileup and subtraction.}
  \label{fig:stamp-pileup05}
\end{figure}

\begin{figure}[p]
  \centering
  \includegraphics[width=0.8\textwidth]{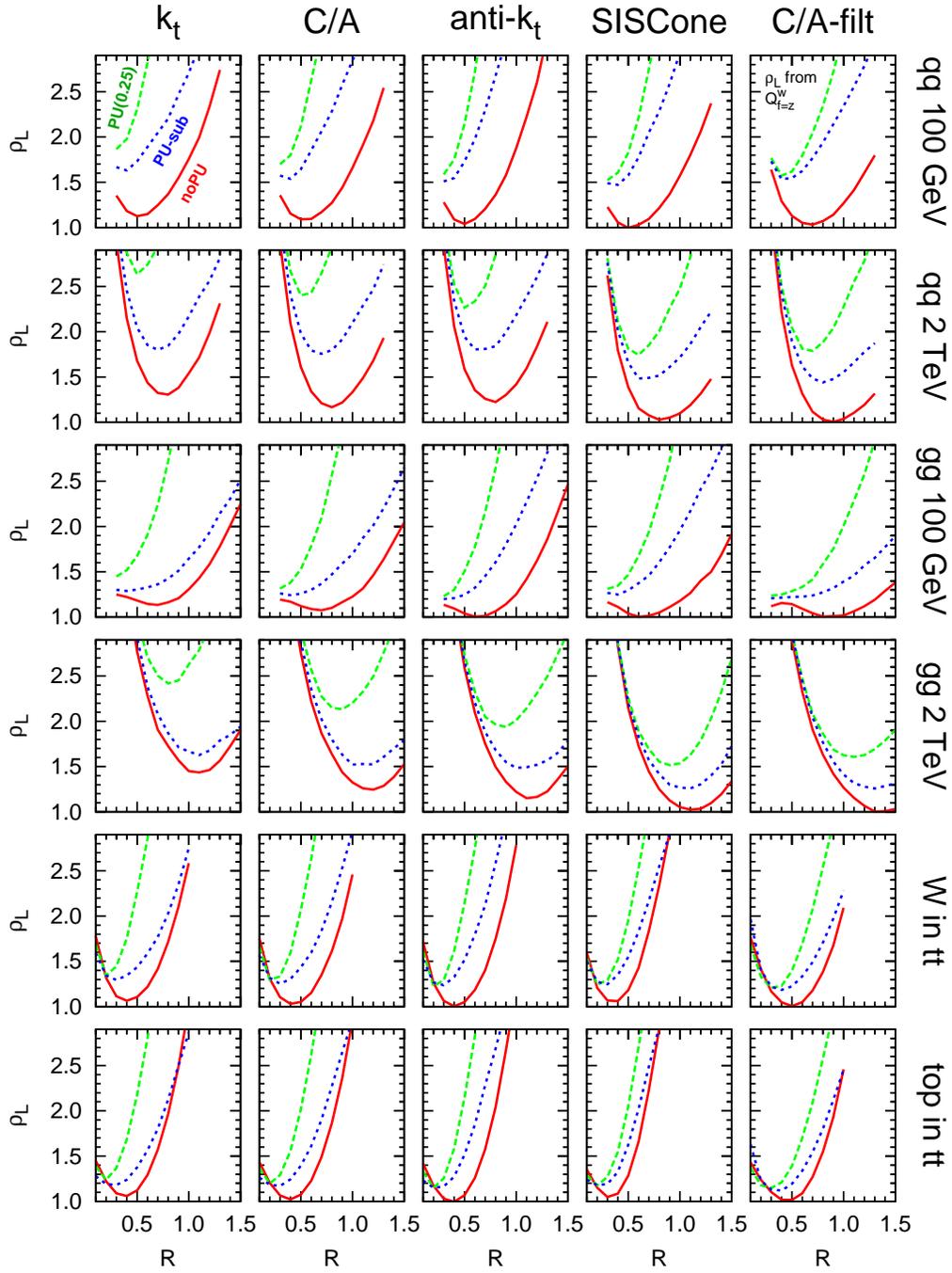}
  \caption{As in figure~\ref{fig:stamp-pileup05}, but for high
    luminosity, $0.25\mb^{-1}$ per bunch crossing.}
  \label{fig:stamp-pileup25}
\end{figure}

Next, in figures~\ref{fig:stamp-pileup05}
and~\ref{fig:stamp-pileup25}, we show results respectively for low and high
pileup. We maintain the same normalisation for the effective
luminosity as before, 
and the (red) solid curves remain those of figure~\ref{fig:summary-no-PU}
throughout (no pileup, no subtraction).
The (green) dashed curves show the effective luminosity ratios with
unsubtracted pileup, while the (blue) dotted curves correspond to
subtracted pileup.
Unsubtracted pileup, unsurprisingly, degrades the quality in almost
all cases, more so at larger $R$ values. This $R$-dependence of the
quality degradation causes the minima to shift to moderately smaller
$R$ values.
Subtraction compensates for part of the loss of quality (and the shift in
best $R$) due to pileup, though to an extent that varies according to
the case at hand.

For the purpose of this article, the main conclusion from
figures~\ref{fig:stamp-pileup05} and~\ref{fig:stamp-pileup25} is that,
if one chooses a jet-definition that is optimal in the case without
pileup (\ie based on the
results of section~\ref{sec:results}), then in the presence of pileup
(with subtraction) 
it gives a $\rhoL$ that remains within about
$10\%$ of the lowest possible value.
Insofar as a given analysis may involve data taken at a range of
instantaneous luminosities (which may even vary significantly over the lifetime of
the beam), this is important since it implies that it will be
satisfactory to choose a single jet-definition independently of the
level of pileup.

\clearpage

\section{Conclusions}\label{sec:conclusions}

In this paper, we have examined the question of assessing the relative
quality of a range of jet definitions for kinematic reconstructions. 
In contrast to other common approaches, we chose not to determine the
quality of a jet definition in terms of how well its jets correspond
to a given (but ill-defined) set of hard Monte-Carlo partons.
Instead we used physically well-defined measures, \ie the
reconstruction of an invariant (dijet or top) mass peak. 
The quality of a given jet-definition is then related to the
``sharpness'' of the mass peak.
Since sharpness is a somewhat fuzzy concept we introduced two
``quality measures'' to quantify the concept, independently of the
peak shape, which is usually strongly non-Gaussian.
With certain hypotheses, one can establish a proportionality relation
between these quality measures and the amount of integrated luminosity
needed to obtain a given statistical significance in a search.

We studied the cases of narrow \qq and \gg resonances over
mass scales ranging from $100\GeV$ to $4\TeV$, as well as top and $W$
reconstruction in $t\bar t$ events. We considered 5 jet algorithms,
$k_t$, Cambridge/Aachen (C/A), anti-$k_t$, SISCone\footnote{
  We recall that anti-$k_t$
  is expected to behave similarly to iterative cones with progressive
  removal (like the current CMS iterative cone), while being infrared
  and collinear (IRC) safe. SISCone is an IRC safe cone algorithm with
  a split--merge step, and is therefore closer to iterative cone
  algorithms like the current ATLAS cone.}
and C/A with filtering, over jet-radii, $R$, spanning typically from
$0.3$ to $1.3$.

Our results (available in a more extensive form via the online
tool~\cite{quality.fastjet.fr})
tend to validate the existing widespread use of low radii
$0.4-0.5$ in reconstructing quark-induced jets at mass scales of
$\order{100\GeV}$. However they also show that gluon-induced jets and
high-scale jets prefer significantly larger $R$, up to an optimal
value of $R \sim 1.2$ for the high-mass \gg case. This general
pattern coincides broadly with analytical expectations~\cite{Dasgupta:2007wa},
and relates to an interplay between needing to capture perturbative
radiation from the jet, while excluding underlying-event
contamination.  The former matters more at high scales and for gluon
jets, hence the preference for larger $R$.

A second pattern that emerges, relevant mainly
at higher energy scales,
is that among traditional (or ``second
generation,'' \cnf footnote~\ref{foot:generations}) types of jet
algorithm, SISCone often performs best and anti-$k_t$ performs better
than other sequential recombination algorithms, $k_t$ and C/A. 
The third-generation C/A-filtering algorithm typically performs as
well as SISCone, but prefers slightly larger $R$. Both SISCone and
C/A-filtering's good performance can be traced back to their low
sensitivity to underlying-event activity.

A quantitative presentation of these results is given in
figure~\ref{fig:summary-no-PU}, for a subset of processes, in terms of
the extra factor in integrated luminosity that would be needed for a
given jet definition to achieve the same significance as the optimal
one in our set.
An implication for LHC experiments planning to use $R=0.4-0.6$ even
in large-mass searches (see \eg\ \cite{Ellis:2007ib,Bhatti:2008hz}), is that some
discoveries may then require up to twice more integrated luminosity
than would be the case with the optimal choice of jet-definition.

Given 
such a statement, it is important
to establish how it is affected by pileup. This was the subject of
section~\ref{sec:results-pu}. The conclusions are that the optimal jet
definition without pileup remains close to optimal even with
high-luminosity pileup (\ie\ with $\sim 25$ $pp$ interactions per bunch
crossings), provided that adequate subtraction methods are used to
correct the jets for the pileup.
Subtraction also reduces the differences between jet algorithms, even
in the absence of pileup, \cnf figure~\ref{fig:stamp-noPUsub-reduced}.

Finally, given that the bulk of our results apply to dijet events, one
may ask to what extent they hold in multi-jet situations.
To investigate this, we have studied hadronic $t\bar t$ events and
observed results similar to those for the low-mass \qq
case. 
However, we envisage that in multi-jet events at high mass scales there
will be 
an additional tension
between the need to resolve the separate jets and the need to include
the bulk of perturbative gluon emission in the jet. 
The study of the issue is beyond the scope of this
article, but we foresee that future developments in third-generation
jet-methods can play an 
important role in optimising analyses of such events.


\section*{Acknowledgements}
\label{sec:acknolwedgments}
This work has been supported in part by the grant ANR-05-JCJC-0046-01
from the French Agence Nationale de la Recherche and under Contract No.
DE-AC02-98CH10886 with the U.S. Department of Energy, and was started
at 2007 Physics at TeV Colliders Les Houches workshop.
It is a pleasure to acknowledge interesting discussions with A.~Oehler and
K.~Rabbertz, we are grateful to G.~Dissertori and G.~Zanderighi for
comments on the
manuscript, and JR thanks D.~d'Enterria for a useful conversation.
GPS wishes to thank Brookhaven National Laboratory and Princeton
University for hospitality while this work was being completed.


\bibliography{jet_algs}

\providecommand{\href}[2]{#2}\begingroup\raggedright\begin{thebibliography}{10}

\bibitem{Buttar:2008jx}
C.~Buttar {\em et.~al.}, {\it {Standard Model Handles and Candles Working
  Group: Tools and Jets Summary Report}},
  \href{http://arXiv.org/abs/0803.0678}{{\tt arXiv:0803.0678}}.

\bibitem{Frixione:2002ik}
S.~Frixione and B.~R. Webber, {\it {Matching NLO QCD computations and parton
  shower simulations}},  {\em JHEP} {\bf 06} (2002) 029,
  [\href{http://arXiv.org/abs/hep-ph/0204244}{{\tt hep-ph/0204244}}].

\bibitem{Nason:2004rx}
P.~Nason, {\it {A new method for combining NLO QCD with shower Monte Carlo
  algorithms}},  {\em JHEP} {\bf 11} (2004) 040,
  [\href{http://arXiv.org/abs/hep-ph/0409146}{{\tt hep-ph/0409146}}].

\bibitem{fit-gaussian-1}
A.~Santocchia, {\it {Optimization of Jet Reconstruction Settings and
  Parton-Level Correction for the $ttH$ Channel}},  tech. rep., 2006.
\newblock CERN-CMS-NOTE-2006-059.

\bibitem{fit-gaussian-2}
A.~P. Cheplakov and S.~Thompson, {\it {MidPoint Algorithm for Jets
  Reconstruction in ATLAS Experiment}},  tech. rep., 2007.
\newblock ATL-PHYS-PUB-2007-007.

\bibitem{quality.fastjet.fr}

\newblock M.~Cacciari, J.~Rojo, G.~P. Salam, and G.~Soyez$,
  \:$\url{http://quality.fastjet.fr/}~.

\bibitem{Sjostrand:2006za}
T.~Sj{\"o}strand, S.~Mrenna, and P.~Skands, {\it Pythia 6.4 physics and
  manual},  {\em JHEP} {\bf 05} (2006) 026,
  [\href{http://arXiv.org/abs/hep-ph/0603175}{{\tt hep-ph/0603175}}].

\bibitem{Albrow:2006rt}
M.~G. Albrow {\em et.~al.}, {\it {Tevatron-for-LHC report of the QCD working
  group}},  \href{http://arXiv.org/abs/hep-ph/0610012}{{\tt hep-ph/0610012}}.

\bibitem{Catani:1991hj}
S.~Catani, Y.~L. Dokshitzer, M.~Olsson, G.~Turnock, and B.~R. Webber, {\it New
  clustering algorithm for multi - jet cross-sections in $e^+ e^-$
  annihilation},  {\em Phys. Lett.} {\bf B269} (1991) 432--438.

\bibitem{Catani:1993hr}
S.~Catani, Y.~L. Dokshitzer, M.~H. Seymour, and B.~R. Webber, {\it
  {Longitudinally invariant $k_t$ clustering algorithms for hadron hadron
  collisions}},  {\em Nucl. Phys.} {\bf B406} (1993) 187--224.

\bibitem{Ellis:1993tq}
S.~D. Ellis and D.~E. Soper, {\it Successive combination jet algorithm for
  hadron collisions},  {\em Phys. Rev.} {\bf D48} (1993) 3160--3166,
  [\href{http://arXiv.org/abs/hep-ph/9305266}{{\tt hep-ph/9305266}}].

\bibitem{Dokshitzer:1997in}
Y.~L. Dokshitzer, G.~D. Leder, S.~Moretti, and B.~R. Webber, {\it Better jet
  clustering algorithms},  {\em JHEP} {\bf 08} (1997) 001,
  [\href{http://arXiv.org/abs/hep-ph/9707323}{{\tt hep-ph/9707323}}].

\bibitem{Wobisch:1998wt}
M.~Wobisch and T.~Wengler, {\it Hadronization corrections to jet cross sections
  in deep- inelastic scattering},
  \href{http://arXiv.org/abs/hep-ph/9907280}{{\tt hep-ph/9907280}}.

\bibitem{Cacciari:2008gp}
M.~Cacciari, G.~P. Salam, and G.~Soyez, {\it {The anti-$k_t$ jet clustering
  algorithm}},  {\em JHEP} {\bf 04} (2008) 063,
  [\href{http://arXiv.org/abs/0802.1189}{{\tt arXiv:0802.1189}}].

\bibitem{Salam:2007xv}
G.~P. Salam and G.~Soyez, {\it A practical seedless infrared-safe cone jet
  algorithm},  {\em JHEP} {\bf 05} (2007) 086,
  [\href{http://arXiv.org/abs/0704.0292}{{\tt arXiv:0704.0292}}].

\bibitem{Cacciari:2008gn}
M.~Cacciari, G.~P. Salam, and G.~Soyez, {\it {The Catchment Area of Jets}},
  {\em JHEP} {\bf 04} (2008) 005, [\href{http://arXiv.org/abs/0802.1188}{{\tt
  arXiv:0802.1188}}].

\bibitem{Butterworth:2008iy}
J.~M. Butterworth, A.~R. Davison, M.~Rubin, and G.~P. Salam, {\it {Jet
  substructure as a new Higgs search channel at the LHC}},  {\em Phys. Rev.
  Lett.} {\bf 100} (2008) 242001, [\href{http://arXiv.org/abs/0802.2470}{{\tt
  arXiv:0802.2470}}].

\bibitem{Cacciari:2005hq}
M.~Cacciari and G.~P. Salam, {\it Dispelling the {$N^3$} myth for the $k_t$
  jet-finder},  {\em Phys. Lett.} {\bf B641} (2006) 57--61,
  [\href{http://arXiv.org/abs/hep-ph/0512210}{{\tt hep-ph/0512210}}].

\bibitem{fastjet_web}

\newblock M.~Cacciari, G.~P. Salam, and G.~Soyez$,
  \:$\url{http://www.fastjet.fr/}~.

\bibitem{Dasgupta:2007wa}
M.~Dasgupta, L.~Magnea, and G.~P. Salam, {\it {Non-perturbative QCD effects in
  jets at hadron colliders}},  {\em JHEP} {\bf 02} (2008) 055,
  [\href{http://arXiv.org/abs/0712.3014}{{\tt arXiv:0712.3014}}].

\bibitem{Ellis:2007ib}
S.~D. Ellis, J.~Huston, K.~Hatakeyama, P.~Loch, and M.~Tonnesmann, {\it {Jets
  in Hadron-Hadron Collisions}},  {\em Prog. Part. Nucl. Phys.} {\bf 60} (2008)
  484--551, [\href{http://arXiv.org/abs/0712.2447}{{\tt arXiv:0712.2447}}].

\bibitem{Bhatti:2008hz}
A.~Bhatti {\em et.~al.}, {\it {CMS search plans and sensitivity to new physics
  with dijets}},  \href{http://arXiv.org/abs/0807.4961}{{\tt arXiv:0807.4961}}.

\bibitem{ORprivate}
{One of us (GPS) wishes to thank A.~Oehler and K.~Rabbertz for a discussion
  related to this point}.

\bibitem{Butterworth:2002tt}
J.~M. Butterworth, B.~E. Cox, and J.~R. Forshaw, {\it {W W scattering at the
  LHC}},  {\em Phys. Rev.} {\bf D65} (2002) 096014,
  [\href{http://arXiv.org/abs/hep-ph/0201098}{{\tt hep-ph/0201098}}].

\bibitem{Seymour:2006vv}
M.~H. Seymour and C.~Tevlin, {\it {A comparison of two different jet algorithms
  for the top mass reconstruction at the LHC}},  {\em JHEP} {\bf 11} (2006)
  052, [\href{http://arXiv.org/abs/hep-ph/0609100}{{\tt hep-ph/0609100}}].

\bibitem{Butterworth:2007ke}
J.~M. Butterworth, J.~R. Ellis, and A.~R. Raklev, {\it Reconstructing sparticle
  mass spectra using hadronic decays},  {\em JHEP} {\bf 05} (2007) 033,
  [\href{http://arXiv.org/abs/hep-ph/0702150}{{\tt hep-ph/0702150}}].

\bibitem{Thaler:2008ju}
J.~Thaler and L.-T. Wang, {\it {Strategies to Identify Boosted Tops}},  {\em
  JHEP} {\bf 07} (2008) 092, [\href{http://arXiv.org/abs/0806.0023}{{\tt
  arXiv:0806.0023}}].

\bibitem{Kaplan:2008ie}
D.~E. Kaplan, K.~Rehermann, M.~D. Schwartz, and B.~Tweedie, {\it {Top-tagging:
  A Method for Identifying Boosted Hadronic Tops}},
  \href{http://arXiv.org/abs/0806.0848}{{\tt arXiv:0806.0848}}.

\bibitem{Almeida:2008yp}
L.~G. Almeida {\em et.~al.}, {\it {Substructure of high-$p_T$ Jets at the
  LHC}},  \href{http://arXiv.org/abs/0807.0234}{{\tt arXiv:0807.0234}}.

\bibitem{Agashe:2006hk}
K.~Agashe, A.~Belyaev, T.~Krupovnickas, G.~Perez, and J.~Virzi, {\it {LHC
  signals from warped extra dimensions}},  {\em Phys. Rev.} {\bf D77} (2008)
  015003, [\href{http://arXiv.org/abs/hep-ph/0612015}{{\tt hep-ph/0612015}}].

\bibitem{Lillie:2007yh}
B.~Lillie, L.~Randall, and L.-T. Wang, {\it {The Bulk RS KK-gluon at the LHC}},
   {\em JHEP} {\bf 09} (2007) 074,
  [\href{http://arXiv.org/abs/hep-ph/0701166}{{\tt hep-ph/0701166}}].

\bibitem{Baur:2008uv}
U.~Baur and L.~H. Orr, {\it {Searching for t-bar t Resonances at the Large
  Hadron Collider}},  {\em Phys. Rev.} {\bf D77} (2008) 114001,
  [\href{http://arXiv.org/abs/0803.1160}{{\tt arXiv:0803.1160}}].

\bibitem{BroojimansTop}
G.~Broojimans, High $p_t$ hadronic top quark identification part {I}: Jet mass
  and ysplitter. ATL-PHYS-CONF-2008-08.

\bibitem{Almeida:2008tp}
L.~G. Almeida, S.~J. Lee, G.~Perez, I.~Sung, and J.~Virzi, {\it {Top Jets at
  the LHC}},  \href{http://arXiv.org/abs/0810.0934}{{\tt arXiv:0810.0934}}.

\bibitem{sLHC}
\url{http://care-hhh.web.cern.ch/CARE-HHH/upgradetable.pdf}~.

\bibitem{Cacciari:2007fd}
M.~Cacciari and G.~P. Salam, {\it {Pileup subtraction using jet areas}},  {\em
  Phys. Lett.} {\bf B659} (2008) 119--126,
  [\href{http://arXiv.org/abs/0707.1378}{{\tt arXiv:0707.1378}}].

\end{thebibliography}\endgroup

\end{document}